\documentclass[final,3p,times,twoside]{elsarticle}

\usepackage{amssymb}
\usepackage{amsmath}
\usepackage{amsthm}
\usepackage{natbib}
\usepackage{amssymb}
\usepackage{mathtools}
\usepackage{multirow}
\usepackage{bm}
\usepackage{stmaryrd}
\usepackage{tabularx}
\usepackage{algorithm}
\usepackage{cancel}
\usepackage{algpseudocode}
\usepackage{tikz}
\usepackage{array}
\usepackage{textcomp}
\usepackage{multirow}
\usepackage{hyperref}
\usepackage{xcolor}
\usepackage{lineno}
\usepackage{eso-pic}
\usepackage{lipsum}

\journal{Elsevier}

\begin{document}

\begin{frontmatter}

\title{Large language model-empowered next-generation computer-aided engineering
}



\author[a]{Jiachen Guo}
\author[b,c]{Chanwook Park}
\author[c,d]{Dong Qian}
\author[e]{Thomas J.R. Hughes}
\author[b,c]{Wing Kam Liu}

\affiliation[a]{organization={Theoretical and Applied Mechanics Program, Northwestern University},
            addressline={2145 Sheridan Road}, 
            city={Evanston},
            postcode={60201}, 
            state={IL},
            country={USA}}

\affiliation[b]{organization={Department of Mechanical Engineering, Northwestern University},
            addressline={2145 Sheridan Road}, 
            city={Evanston},
            postcode={60201}, 
            state={IL},
            country={USA}}

\affiliation[c]{organization={HIDENN-AI, LLC},
            addressline={1801 Maple Ave}, 
            city={Evanston},
            postcode={60201}, 
            state={IL},
            country={USA}}

\affiliation[d]{organization={Department of Mechanical Engineering, University of Texas, Dallas},
            addressline={800 W. Campbell Road}, 
            city={Richardson},
            postcode={75080}, 
            state={TX},
            country={USA}}

\affiliation[e]{organization={Oden Institute for Computational Engineering and Sciences, University of Texas at Austin},
            addressline={201 E 24th St}, 
            city={Austin},
            postcode={78712}, 
            state={TX},
            country={USA}}

\begin{abstract}

Software development has entered a new era where large language models (LLMs) now serve as general-purpose reasoning engines, enabling natural language interaction and transformative applications across diverse domains. This paradigm is now extending into computer-aided engineering (CAE), offering a potential solution to the significant human effort and computational expense that constrain traditional finite element methods (FEM), particularly for large-scale parametric problems. Recent applications of LLMs in CAE have successfully automated routine tasks, including CAD model generation and FEM simulations. Nevertheless, these contributions, which primarily serve to reduce manual labor, are often insufficient for addressing the significant computational challenges posed by large-scale, high-dimensional systems. To this aim, we first introduce the concept of LLM-empowered CAE agent, where LLMs act as autonomous collaborators that plan, execute, and adapt CAE workflows. Then, we propose LLM-empowered CAE agent for data-free model order reduction (MOR), a powerful yet underused approach for ultra-fast large-scale parametric analysis due to the intrusive nature and labor-intensive redevelopment of solvers. LLMs can alleviate this barrier by automating derivations, code restructuring, and implementation, making intrusive MOR both practical and broadly accessible. To demonstrate feasibility, we present an LLM-empowered CAE agent for solving ultra-large-scale space-parameter-time (S-P-T) physical problems using Tensor-decomposition-based A Priori Surrogates (TAPS). Our results show that natural language prompts describing parametric partial differential equations (PDEs) can be translated into efficient solver implementations, substantially reducing human effort while producing high-fidelity reduced-order models. Moreover, LLMs can synthesize novel MOR solvers for unseen cases such as nonlinear and high-dimensional parametric problems based on their internal knowledge base. This highlights the potential of LLMs to establish the foundation for next-generation CAE systems.

\end{abstract}




\end{frontmatter}
\textbf{Highlight}
\begin{itemize}   


\item \textbf{CAE agents for autonomous CAE workflows:} LLM can act as proactive collaborators that plan, execute and adapt CAE workflows to save human labor and, more importantly, can learn from a growing knowledge database.

\item \textbf{Overcoming barriers in intrusive model order reduction (MOR):} LLMs offer a path to automate the algebraic derivations and code restructuring required for data-free intrusive MOR, making this powerful but underused method more practical and accessible to CAE engineers.

\item \textbf{Reducing LLM hallucination in data-Free MOR:} We show that a Chain-of-Thought prompting strategy, guided by curated examples, significantly reduces LLM hallucinations when deriving the complex mathematical equations required for intrusive, data-free MOR solvers.

\item \textbf{Extrapolating solver capabilities beyond simple examples:} LLMs can develop sophisticated data-free MOR solvers (i.e., nonlinear/high-dimensional parametric problems) by correctly generalizing from provided simpler linear examples, autonomously integrating advanced numerical schemes from their internal knowledge base.

\item \textbf{Toward next-generation CAE systems:} LLM-empowered CAE agents can deliver surrogate models that achieve accuracy, speed, scalability and efficiency simultaneously, paving the way for future CAE platforms for engineering design and optimization.

\end{itemize}

\section{Introduction}
\label{sec1}

Software development is undergoing a paradigm shift, transitioning from explicit human-coded algorithms to systems defined by pre-trained neural networks \cite{Karpathy_2017}. In the latest phase of this evolution, the natural language itself serves as the programming interface, while large language models (LLMs) perform the underlying reasoning and implementation \cite{Karpathy_2025}. Within this new paradigm, developers leverage LLMs as general-purpose reasoning engines, defining objectives in natural language to generate the required implementation \cite{raiaan2024review} automatically. The impact of this shift extends far beyond computer science: LLMs are already driving advances in biomedical research and healthcare \cite{nazi2024large}, transforming educational tools and practices \cite{kasneci2023chatgpt, jeon2023large}, and augmenting expert work in high-stakes fields such as finance, consulting, and law \cite{chen2024survey}. Collectively, these trends signal a fundamental transformation in the way we harness computation across diverse domains.

The paradigm shift can also be projected to the domain of computer-aided engineering (CAE). Standard implementations of the finite element method (FEM), on commercial or open-source platforms, exemplify the traditional software paradigm. These frameworks rely on fixed, hard-coded numerical procedures to solve the governing equations for physical phenomena such as structural mechanics, heat transfer, and fluid dynamics \cite{liu2022eighty}. The first wave of machine learning integration focused on improving these core numerical methods, using the universal approximation capability of neural networks to improve the accuracy and efficiency of FEM through adaptive meshing and basis functions \cite{zhang2021hierarchical, zhang2022hidenn, lu2023convolution, park2023convolution, li2023convolution, park2024interpolating, guo2025interpolation, guo2025tensor}. More recently, a second wave, driven by large language models (LLMs), is automating higher-level, labor-intensive processes, such as the creation of computer-aided design (CAD) models in standard CAE workflow \cite{govindarajan2025cadmium, lv2025cadinstruct, zhang2025large}. In addition, recent attempts have focused on automating FEM simulations with legacy solvers. Hou et al. proposed AutoFEA, which reduces AI hallucinations by using a graph convolutional network transformer link prediction retrieval model \cite{hou2025autofea}; Mudur et al. benchmarked the success rate of LLM agents in conjunction with COMSOL Multiphysics\textsuperscript{\textregistered} \cite{mudur2024feabench}; Pandey et al. extended LLM capabilities through retrieval-augmented generation (RAG) for running simulations in OpenFOAM \cite{pandey2025openfoamgpt}. However, most of these applications only use LLMs as a function-calling tool to invoke standard CAE software, a task readily performed by human experts. In practice, this means that if a simulation fails to converge or produces physically unrealistic results, the system cannot independently diagnose the problem. For example, it does not know how to refine the mesh near a stress concentration or adjust the timestep to resolve a contact issue. 



A CAE agent is an LLM-empowered system that acts as a proactive engineering collaborator, moving beyond the passive execution of predefined scripts. By leveraging an expandable ``toolbox" of specialized functions described in natural language, CAE agents can autonomously plan and adapt entire analysis workflows, including designing experiments, running solvers, monitoring convergence, refining meshes, and post-processing. This allows it to intelligently automate the complex, trial-and-error decisions inherent in CAE, such as choosing a suitable material model or refining a mesh based on intermediate results. Therefore, CAE agents can capture and scale expert knowledge to greatly accelerate the design cycle by saving human labor.

Although CAE agents equipped with standard numerical methods, such as FEM, excel in optimizing the simulation workflow, a more fundamental challenge is the computational cost of the solvers themselves, especially for large-scale problems. This is where Model Order Reduction (MOR) tools offer another transformative solution. MOR techniques are typically classified into two categories: data-driven methods, including neural operators and other neural network-based surrogates \cite{kovachki2023neural, lehtimaki2022accelerating, park2024interpolating, guo2025interpolation, mozaffar2019deep}. Due to their non-intrusive nature, they can be applied broadly without modifying the original legacy solvers. However, they often require large amounts of offline simulation data as training data, which is prohibitive to generate for large-scale, high-dimensional parametric problems. In contrast, data-free MOR methods, including proper generalized decomposition \cite{chinesta2013proper}and the more recent Hierarchical Deep-learning Neural Network Tensor Decomposition (HiDeNN-TD) with its variants \cite{zhang2022hidenn, lu2023convolution, li2023convolution, guo2024convolutional}, involve direct modifications to the solver's core algorithms and are intrusive by nature. Although these methods do not require training data and offer significant efficiency for extremely large-scale problems, their adoption is severely hindered by the effort required for extensive redevelopment of existing legacy codes, which poses a substantial barrier for most CAE engineers.

Recent breakthroughs in the mathematical reasoning and code generation abilities of LLMs offer a direct solution to this implementation barrier. State-of-the-art models have demonstrated expert-level performance in these domains \cite{huang2025gemini, chervonyi2025gold, shalyt2025asymob}, suggesting that they can automate the most labor-intensive aspects of data-free intrusive MOR, from initial algebraic derivations to complex restructuring and implementation of solver code. This represents a fundamental leap beyond previous standard CAE agents, which focused on automating the workflow using existing legacy solvers \cite{govindarajan2025cadmium, lv2025cadinstruct, zhang2025large, hou2025autofea, mudur2024feabench, pandey2025openfoamgpt}. In contrast, our proposed approach targets the core scientific challenge itself: the complex re-engineering of the solver architecture. Leveraging advanced, pre-trained language models thus provides a pathway to automate solver development, unlocking the full potential of data-free MOR with unprecedented efficiency while ensuring accuracy.

\begin{figure}[!hbt]
\centering
\includegraphics[width=0.75\linewidth]{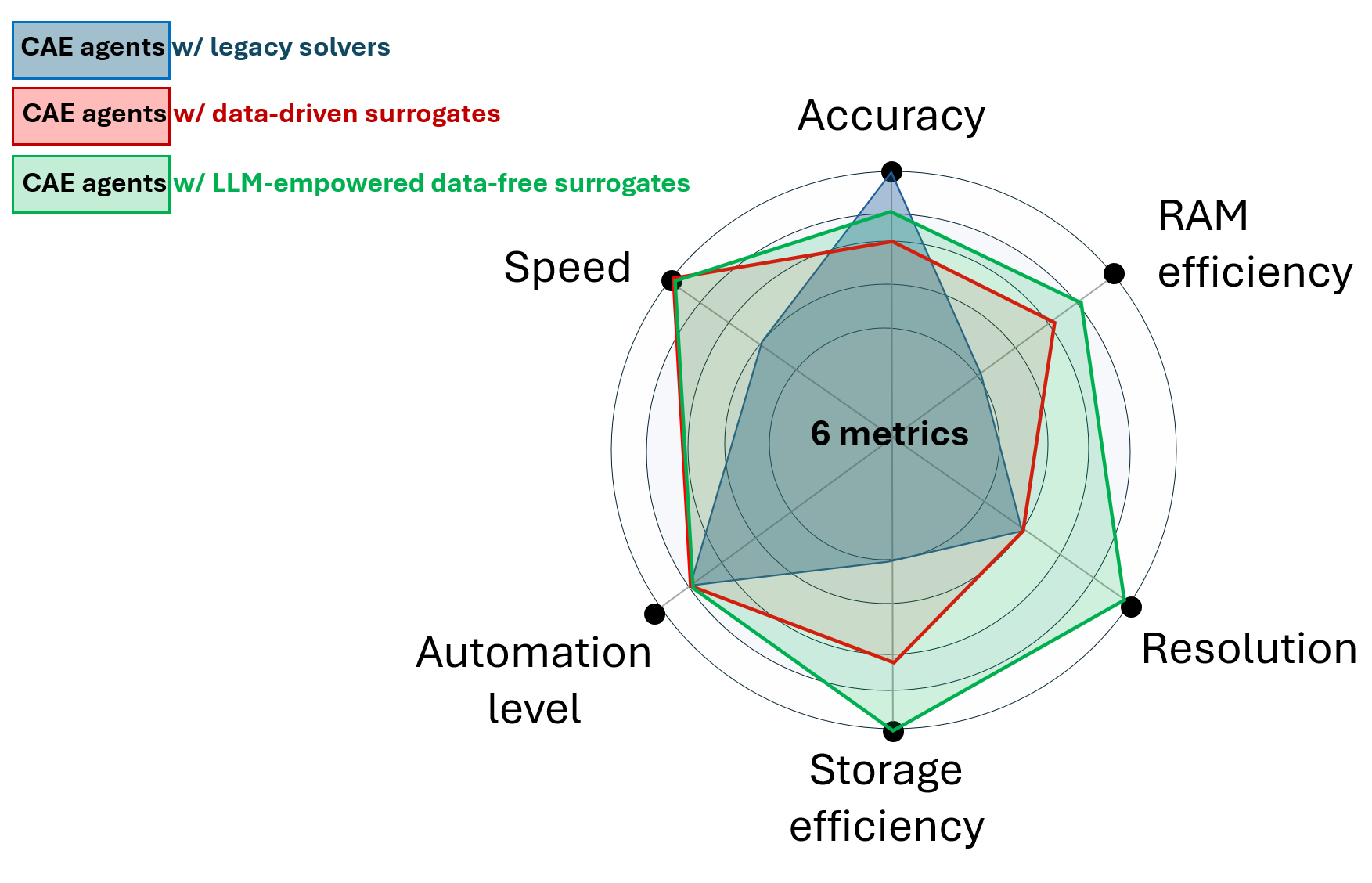}
\caption{Comparison of different CAE agents with different tools in 6 performance metrics. CAE agents with legacy solvers have a high automation level and good accuracy, but suffer from computational cost for large-scale problems. CAE agents with data-driven surrogates can make predictions very fast once well trained, but the prediction accuracy is relatively limited. Training data-driven models can also be costly for large-scale, high-dimensional problems. CAE agents with LLM-empowered data-free MOR solvers, however, don't require any offline training data. As a result, it has faster solving speed, better RAM and storage efficiency, and can easily handle large-scale problems that require a high-resolution mesh.}
\label{fig:metric6}
\end{figure}

In this article, we introduce a novel CAE agent designed to automate the development of data-free intrusive MOR. Specifically, we focus on Tensor-decomposition-based A Priori Surrogates (TAPS) as the adopted data-free model \cite{guo2025tensor} and show how natural language prompts can be translated into efficient solver implementations. The proposed approach significantly reduces human effort in the development of intrusive MOR models, including mathematical derivations, code implementation, and verification studies. As shown in Fig. \ref{fig:metric6}, the resulting approach supersedes other tools across multiple performance metrics for large-scale, high-dimensional parametric problems. Through numerical examples, we illustrate the feasibility of this approach and discuss its potential as a foundation for future CAE systems that simultaneously meet the key performance requirements of accuracy, speed, resolution, memory efficiency, and scalability. This paper is structured as follows. First, we present the general concept of a CAE agent in Section \ref{sec:Agentic_AI}. In Section \ref{sec:MOR}, we introduce the basics of the adopted data-free intrusive MOR, i.e., Tensor-decomposition-based A Priori Surrogates (TAPS), and how LLMs can be leveraged to automate the development of TAPS for different parametric PDEs while overcoming LLMs' hallucinations. Then we present the capability of the proposed framework via a variety of numerical examples in Section \ref{subsec:results}. Finally, in Sections \ref{sec:discussion} and \ref{sec:conclusion}, we discuss the potential improvements of the current approach in the near future.

\section{Preliminary of CAE agent}
\label{sec:Agentic_AI}


\begin{figure}[!hbt]
\centering
\includegraphics[width=1.0\linewidth]{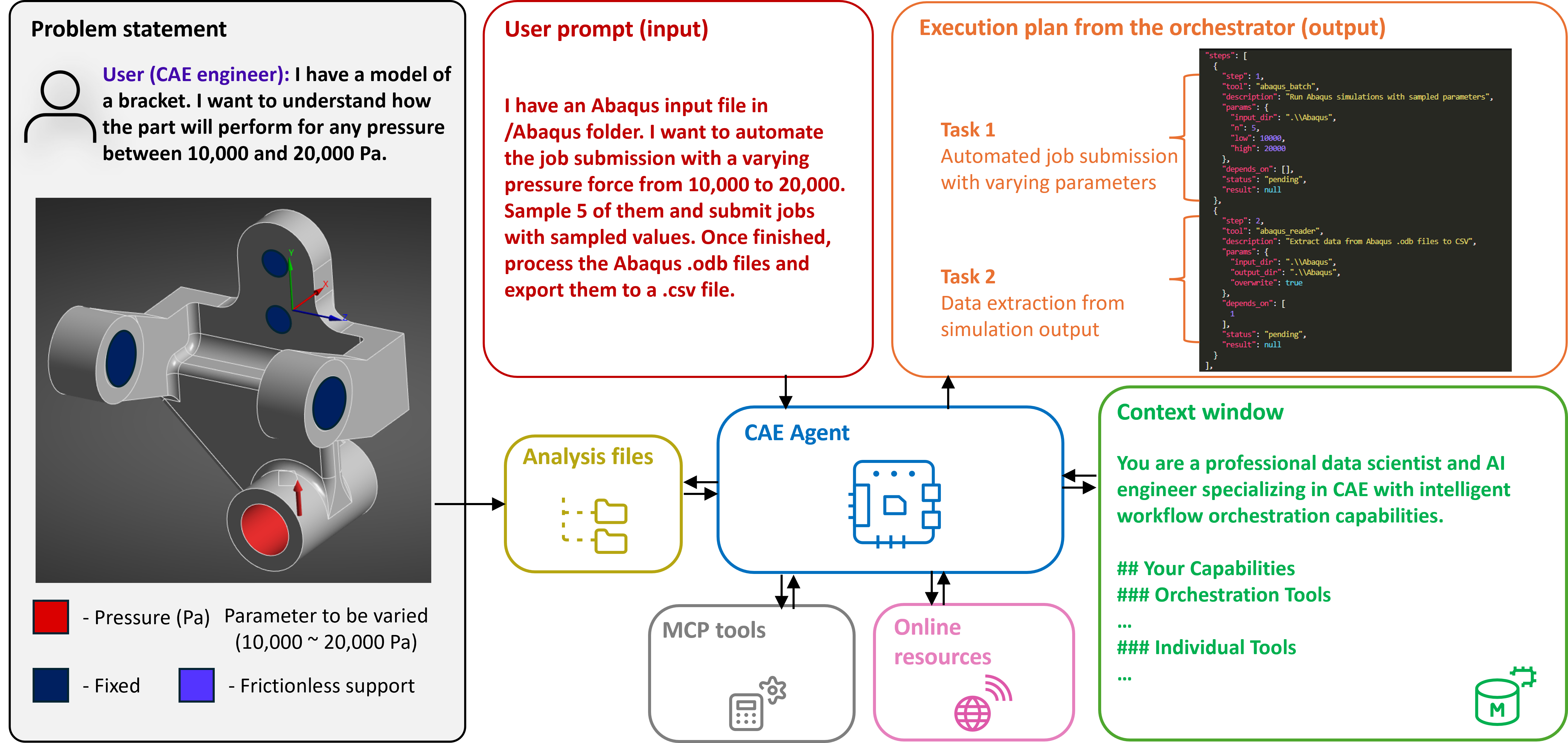}
\caption{CAE agent for a data-driven surrogate modeling. The problem statement is provided in the left box. The user prompt describes the task more technically, and the execution plan from the orchestrator deciphers the user prompt, converting it into a series of executable model context protocol (MCP) tools. }
\label{fig:cae_agent}
\end{figure}

In this section, we first introduce the general framework of a CAE agent: an autonomous system powered by a Large Language Model (LLM) and equipped with multiple CAE tools. The CAE agent is built by incorporating various modules with an LLM, where each module teaches the domain expertise of CAE to the LLM. As illustrated in Figure \ref{fig:cae_agent}, the CAE agent consists of four key modules: analysis files, tools, online resources, and context window. To illustrate the roles of each module, we introduce a simple data-driven surrogate modeling task in CAE.

The problem statement is described in the left box of Figure \ref{fig:cae_agent}. Consider a user who wants to create a surrogate model of a bracket subjected to a static pressure force acting on the bottom cylinder. This is a data-driven problem that involves a series of CAE tasks, including multiple job submissions with varying parameters, data extraction from the simulation output, data preprocessing, surrogate model training, and visualization. For an LLM to handle this CAE task, we first need to provide a working simulation file. For example, we can provide an input file of a commercial software (e.g., Abaqus) that reflects the problem definition, including the part geometry, boundary conditions, loading conditions, etc. Although we want to parameterize the loading condition (i.e., pressure force), we can leave it as a fixed number because the CAE agent can update it later on.

Another critical component for building a CAE agent is the context window, which defines how the LLM should behave as an engineering assistant. In this setup, the context window begins with the instruction: ``You are a professional data scientist and AI engineer specializing in CAE and intelligent workflow orchestration capabilities”, as illustrated in Figure \ref{fig:cae_agent}. This statement establishes the persona of the LLM as a CAE agent and specifies both the available tools and the general expectations of its role. The tools listed in the context window must correspond to those registered through the model context protocol (MCP) \cite{hou2025model}. MCP serves as a universal adapter that enables LLMs to interact with various applications, tools, and data sources. In this problem setup, the MCP tools include automated job submission with varying parameters, data extraction from simulation output, surrogate model training, and visualization. Each MCP tool is provided with contextual guidance on when and how it should be used and what its capabilities are. A detailed explanation of how these MCP tools are built and connected to an LLM is provided in \ref{Appendix:mcp}. Finally, if relevant online resources are available, such as user documentation for CAE software, they can also be linked to the CAE agent to provide up-to-date references that support decision-making and workflow execution. This can be achieved through retrieval-augmented generation (RAG) \cite{lewis2020retrieval}.

The critical aspect of the CAE agent is that it acts as an autonomous agent capable of planning, executing, and adapting CAE workflows, rather than an automated function-calling tool for each subtask. Therefore, the CAE agent functions more like a proactive collaborator in engineering design and analysis, enabling faster exploration of parameter spaces and more efficient use of computational resources. To achieve this, we need to enhance both the context window and the MCP, allowing the CAE agent to plan and execute a series of tool calls. With this setup, once the user prompt is provided, the CAE agent converts it to an executable work plan that calls the available MCP tools. For example, as illustrated in Figure \ref{fig:cae_agent}, the CAE agent converts the user prompt into an execution plan that calls two MCP tools: automated job submission and data extraction. Once the user approves, it will perform the two tasks automatically. The current demonstration is available for free to interested readers. \footnote{\url{https://hidennai.wordpress.com/}}.

One possible critique of this workflow is that it resembles a conventional automation tool. This is a fair observation when dealing with relatively small problems that can be handled directly by human experts. However, the true transformative potential of the CAE agent lies in its scalability. As the domain expertise provided accumulates (i.e., analysis files, context window, MCP tools, and online resources) and the size of the problem becomes extremely large, the system eventually surpasses the cognitive capacity of a single human expert to fully manage. The development of a new intrusive data-free MOR tool exemplifies this threshold, a case that we will further elaborate on in the following sections of this article.


\section{CAE agent for data-free model order reduction}
\label{sec:MOR}
\subsection{Tensor-decomposition-based A Priori Surrogates (TAPS)}
\subsubsection{Tensor decomposition approximation}

The computational modeling of many physical phenomena is challenged by their inherently multiscale and high-dimensional nature. Additive manufacturing, for instance, presents a formidable challenge to classical simulation tools such as the FEM because of the vast disparity in the scales involved. Resolving the micron-scale dynamics of the melt pool across the centimeter-scale geometry of the final component results in a huge number of degrees of freedom (DoFs) that are computationally intensive. Furthermore, the simulation constitutes a high-dimensional parametric problem, as the thermal field depends on numerous process/material parameters, such as laser spot size, power, scanning speed, laser absorptivity, and material diffusivity, which gives rise to the curse of dimensionality. To address such challenges, we adopt Tensor-decomposition-based A Priori Surrogates (TAPS) as the intrusive data-free MOR model \cite{guo2025tensor}. TAPS is data-free since no training data is required to solve the unknowns in the surrogate model. Instead, we directly plug the approximation into the governing equation and solve the unknowns by minimizing the PDE residual. 

We first introduce the basic formulation of TAPS. Assume a general $D$-dimensional space-parameter-time (S-P-T) problem with independent variables defined in Eq. \ref{variables}:

\begin{equation}
    \bm{x}=(\underbrace{x_1,...,x_{S}}_\textrm{spatial variables}, \underbrace{x_{S+1},...,x_{P}}_\textrm{parametric variables}, x_t)
\label{variables}
\end{equation}
where $\bm{x}_s=(x_1,...,x_{S})$ refer to spatial variables; $\bm{x}_p=(x_{S+1},...,x_{P})$ are parametric variables; $x_t$ represents time.
TAPS leverages tensor decomposition (TD) to approximate the solution field $u(\bm{x}_s, \bm{x}_p, x_t)$ as a sum of multiplication of univariate functions:
\small
\begin{equation}
    u^{TD}(\bm{x}_s, \bm{x}_p,{x}_t)=\sum_{m=1}^M \underbrace{\left[u^{(m)}_1(x_1) \cdots u^{(m)}_S(x_S)\right]}_\textrm{spatial }\underbrace{\left[u^{(m)}_{S+1}(x_{S+1}) \cdots u^{(m)}_P(x_P)\right]}_\textrm{parametric} \underbrace{u^{(m)}_{t}(x_t)}_\textrm{temporal}
\label{td_eq_0}
\end{equation}
\normalsize
where $M$ represents the number of modes in TD.

Eq. \ref{td_eq_0} can be simplified using the following product notation:
\begin{equation}
    u^{TD}(\bm{x}_s, \bm{x}_p,{x}_t)=\sum_{m=1}^M\prod_{d=1}^Du^{(m)}_d(x_d)
\label{td_eq}
\end{equation}
where $u^{(m)}_d(x_d)$ is the univariate function for dimension $d$ and mode $m$; $D$ is the total number of independent variables. 

After decomposing the original high-dimensional multivariate function into a finite sum of multiplications of univariate functions to overcome the curse of dimensionality, each of these functions, $u^{(m)}_d(x_d)$, can then be parameterized using different approximation schemes, such as finite elements, radial basis functions, splines, or multilayer perceptrons (MLPs). In this paper, we use a novel AI-enhanced basis function called Convolution Hierarchical Deep-learning Neural Network (C-HiDeNN). This approach marries the merits of both locally supported finite element shape functions and the flexibility of machine learning. Consequently, C-HiDeNN not only maintains all the essential finite element approximation properties but also has controllable accuracy by virtue of the flexibility of neural networks. The detailed formulation of C-HiDeNN can be found in \ref{A:chidenn}. As a result, the C-HiDeNN-TD approximation of a general S-P-T problem can be written as:

\begin{equation}
    u^{TD}(\bm{x}_s, \bm{x}_p,{x}_t)=\sum_{m=1}^M\prod_{d=1}^D\widetilde{\bm{N}}^{[d]}(x_d)\bm{u}^{[d]}_{m}
\label{td_eq}
\end{equation}
where $\widetilde{\bm{N}}^{[d]}(x_d)$ is the C-HiDeNN shape function for the $d$-th dimension; $\bm{u}^{[d]}_{m}$ refers to the discretized solution vector for $m$-th mode and $d$-th dimension.

\subsubsection{Formulation of TAPS}
As a data-free surrogate modeling approach, TAPS does not require offline training data to obtain the surrogate model. In contrast, TAPS solves the unknowns in the surrogate model $\bm{u}^{[d]}_{m}$ a priori using the generalized Galerkin projection. As a demonstration, we present the mathematical derivation for the following S-P-T PDE. 

\begin{equation}
    \frac{\partial u}{\partial t}-\frac{\partial }{\partial x}\left( \alpha \frac{\partial u}{\partial x}\right)=f(x, \alpha,t)
\label{pde_eqn1}
\end{equation}
which is subject to the homogeneous Dirichlet boundary and initial conditions. This is a parametric PDE that models 1D transient heat transfer. The equation has 3 independent variables ($D=3$), i.e., the spatial variable $x_s = x_1=x$, the parametric variable $x_p = x_2= \alpha$, and the temporal variable $x_t = x_3 =t$. For simplicity, we assume $f(x, \alpha,t) = f_x(x)f_{\alpha}(\alpha)f_t(t)$. The S-P-T weak form of this PDE can be written as:

\begin{equation}
    \int_{\Omega}\delta u \nabla_{t} u dx d\alpha dt +\int_{\Omega}\nabla_{x}\delta u\cdot \alpha\nabla_{x}udx d\alpha dt - \int_{\Omega}\delta u f(x,\alpha, t) dx d\alpha dt = 0
\label{galerkin10}
\end{equation}
where $\Omega$ is the S-P-T continuum. We define the following trial solution using C-HiDeNN-TD approximation. 

\begin{equation}
    u(x,k,t)=\sum_{m=1}^M \left[\widetilde{\bm{N}}^{[x]}(x)\bm{u}_m^{[x]} \right] \left[\widetilde{\bm{N}}^{[\alpha]}(\alpha)\bm{u}_m^{[\alpha]}\right]  \left[\widetilde{\bm{N}}^{[t]}(t)\bm{u}_m^{[t]}\right]
\label{trial}
\end{equation}
The corresponding test function is obtained using the variational principle.

\begin{equation}
\delta u(x,\alpha,t) =
\begin{aligned}[t]
  &\underbrace{\sum_{m=1}^M
   \big[\widetilde{\bm{N}}^{[x]}(x)\delta \bm{u}_m^{[x]}\big]
   \big[\widetilde{\bm{N}}^{[\alpha]}(\alpha)\bm{u}_m^{[\alpha]}\big]
   \big[\widetilde{\bm{N}}^{[t]}(t)\bm{u}_m^{[t]}\big]}_{\text{spatial variation}}
  \\[1ex]
  &+\underbrace{\sum_{m=1}^M
   \big[\widetilde{\bm{N}}^{[x]}(x)\bm{u}_m^{[x]}\big]
   \big[\widetilde{\bm{N}}^{[\alpha]}(\alpha)\delta\bm{u}_m^{[\alpha]}\big]
   \big[\widetilde{\bm{N}}^{[t]}(t)\bm{u}_m^{[t]}\big]}_{\text{parametric variation}}
  \\[1ex]
  &+\underbrace{\sum_{m=1}^M
   \big[\widetilde{\bm{N}}^{[x]}(x)\bm{u}_m^{[x]}\big]
   \big[\widetilde{\bm{N}}^{[\alpha]}(\alpha)\bm{u}_m^{[\alpha]}\big]
   \big[\widetilde{\bm{N}}^{[t]}(t)\delta\bm{u}_m^{[t]}\big]}_{\text{temporal variation}}.
\end{aligned}
\label{test}
\end{equation}
Plugging Eqs. \ref{trial}-\ref{test} into \ref{galerkin10}, we can obtain a nonlinear system of equations with respect to the generalized S-P-T DoFs $\mathbb{U}=\left(\mathbb{U}^{[x]}, \mathbb{U}^{[\alpha]}, \mathbb{U}^{[t]}\right)$, where $\mathbb{U}^{[d]}$ denotes the vectorized solution vector for all modes in the $d$-th dimension $\mathbb{U}^{[d]}= \left[\bm{u}_1^{[d]}, \cdots \bm{u}_m^{[d]}\right]$. The subspace iteration algorithm is then leveraged to approximate the solution to the nonlinear equation by iteratively solving a series of linear problems in the subspace. Each linear subspace problem solves the following system of equations. \ref{app:A} shows details of subspace iteration and how the corresponding matrix form is derived.
\begin{equation}
    \left( \bm{A}^{[d]}  \otimes \bm{K}_1^{[d]} + \bm{B}^{[d]}  \otimes \bm{K}_2^{[d]} \right) \text{vec}(\mathbb{U}^{[d]}) = \text{vec}(\mathbb{Q}^{[d]})
\end{equation}
where $\text{vec}(\cdot)$ refers to vectorization; $\bm{K}_1^{[d]}, \bm{K}_2^{[d]}$ are sparse matrices related to spatial/parametric/temporal solution from the previous iteration; $\bm{A}^{[d]}, \bm{B}^{[d]}$ are coefficient matrices that depend on modal coefficients from other dimensions; $\otimes$ denotes the Kronecker product. The sparse matrices for each dimension are defined as follows:

Spatial matrices in dimension $x$:
\begin{subequations}
\begin{gather}
\bm{K}^{[x]} =\bm{K}^{[B_xB_x]} = \bm{K}^{[x]}_1 = \int_{\Omega_x} \widetilde{\bm{B}}^{[x]T}(x) \widetilde{\bm{B}}^{[x]}(x) \, dx \quad \text{(stiffness matrix)}\\
\bm{M}^{[x]} = \bm{K}^{[N_xN_x]} = \bm{K}^{[x]}_2= \int_{\Omega_x} \widetilde{\bm{N}}^{[x]T}(x) \widetilde{\bm{N}}^{[x]}(x) \, dx \quad \text{(mass matrix)}
\end{gather}
\end{subequations}
where the C-HiDeNN shape function derivative $\widetilde{\bm{B}}^{[x_d]}(x_d) = \frac{d \widetilde{\bm{N}}^{[x_d]}(x_d)}{d x_d}$.

Parametric matrices in dimension $\alpha$:
\begin{subequations}
\begin{gather}
\bm{K}^{[N_\alpha \alpha N_\alpha]} = \bm{K}^{[\alpha]}_1 = \int_{\Omega_\alpha} \widetilde{\bm{N}}^{[\alpha]T}(\alpha) \alpha \widetilde{\bm{N}}^{[\alpha]}(\alpha) \, d\alpha \\
\bm{K}^{[N_\alpha N_\alpha]} = \bm{K}^{[\alpha]}_2 = \int_{\Omega_\alpha} \widetilde{\bm{N}}^{[\alpha]T}(\alpha) \widetilde{\bm{N}}^{[\alpha]}(\alpha) \, d\alpha
\end{gather}
\end{subequations}

Temporal matrices in dimension $t$:
\begin{subequations}
\begin{gather}
\bm{K}^{[N_t N_t]} = \bm{K}^{[t]}_1 = \int_{\Omega_t} \widetilde{\bm{N}}^{[t]T}(t) \widetilde{\bm{N}}^{[t]}(t) \, dt\\
\bm{K}^{[N_t B_t]} = \bm{K}^{[t]}_2 = \int_{\Omega_t} \widetilde{\bm{N}}^{[t]T}(t) \widetilde{\bm{B}}^{[t]}(t) \, dt
\end{gather}
\end{subequations}

The coefficient matrices for each dimension can be derived as:

For dimension $x$:
\begin{subequations}
\begin{gather}
\bm{A}^{[x]} = (\mathbb{U}^{[t]T} \bm{K}^{[N_t N_t]} \mathbb{U}^{[t]}) \odot (\mathbb{U}^{[\alpha]T} \bm{K}^{[N_\alpha \alpha N_\alpha]} \mathbb{U}^{[\alpha]})\\
\bm{B}^{[x]} = (\mathbb{U}^{[t]T} \bm{K}^{[N_t B_t]} \mathbb{U}^{[t]}) \odot (\mathbb{U}^{[\alpha]T} \bm{K}^{[N_\alpha N_\alpha]} \mathbb{U}^{[\alpha]})
\end{gather}
\end{subequations}

where $\odot$ refers to the element-wise product.

For dimension $\alpha$:
\begin{subequations}
\begin{gather}
\bm{A}^{[\alpha]} = (\mathbb{U}^{[x]T} \bm{K}^{[x]} \mathbb{U}^{[x]}) \odot (\mathbb{U}^{[t]T} \bm{K}^{[N_t N_t]} \mathbb{U}^{[t]})\\
\bm{B}^{[\alpha]} = (\mathbb{U}^{[x]T} \bm{M}^{[x]} \mathbb{U}^{[x]}) \odot (\mathbb{U}^{[t]T} \bm{K}^{[N_t B_t]} \mathbb{U}^{[t]})
\end{gather}
\end{subequations}

For dimension $t$:
\begin{subequations}
\begin{gather}
\bm{A}^{[t]} = (\mathbb{U}^{[x]T} \bm{K}^{[x]} \mathbb{U}^{[x]}) \odot (\mathbb{U}^{[\alpha]T} \bm{K}^{[N_\alpha \alpha N_\alpha]} \mathbb{U}^{[\alpha]})\\
\bm{B}^{[t]} = (\mathbb{U}^{[x]T} \bm{M}^{[x]} \mathbb{U}^{[x]}) \odot (\mathbb{U}^{[\alpha]T} \bm{K}^{[N_\alpha N_\alpha]} \mathbb{U}^{[\alpha]})
\end{gather}
\end{subequations}

The index form of the force vector $\bm{Q}^{[d]}$ for each dimension can be derived as:\\

For $x$-direction:
$$\mathbb{Q}^{[x]}_{n_x m} = Q_{n_x}^{[x]} \cdot \left( \sum_{n_\alpha} u_{n_\alpha m}^{[\alpha]} I_{n_\alpha}^{[\alpha]} \right) \cdot \left( \sum_{n_t} u_{n_t m}^{[t]} I_{n_t}^{[t]} \right)$$

For $t$-direction:
$$\mathbb{Q}^{[t]}_{n_t m} = Q_{n_t}^{[t]} \cdot \left( \sum_{n_x} u_{n_x m}^{[x]} I_{n_x}^{[x]} \right) \cdot \left( \sum_{n_\alpha} u_{n_\alpha m}^{[\alpha]} I_{n_\alpha}^{[\alpha]} \right)$$

For $\alpha$-direction:
$$\mathbb{Q}^{[\alpha]}_{n_\alpha m} = Q_{n_\alpha}^{[\alpha]} \cdot \left( \sum_{n_x} u_{n_x m}^{[x]} I_{n_x}^{[x]} \right) \cdot \left( \sum_{n_t} u_{n_t m}^{[t]} I_{n_t}^{[t]} \right)$$

where the subscript $n_d$ refers to the number of discretized grid points in dimension $d$. The integration vectors are defined as:
$$
Q_{n_d}^{[d]} = \int_{\Omega_d} \widetilde{N}_{n_d}^{[d]}(x_d) \, dx_d
$$

$$I_{n_d}^{[d]} = \int_{\Omega_d} \widetilde{N}_{n_d}^{[x_d]}(x_d) f_{x_d}(x_d) \, dx_d$$

As can be seen above, the final matrix form of the linear system in each subspace iteration is complicated. Deriving these matrices from the original S-P-T PDE presents the following challenges: 

(1) \textbf{Complex Tensor Algebra}:
Deriving the matrix form from the S-P-T weak form requires complex tensor algebra, including operations such as contraction, vectorization, and the Kronecker product. Although symbolic programming is useful for analytical derivations, its application to complex tensor algebra, particularly for high-order tensors, remains a significant challenge.

(2) \textbf{Susceptibility to Human Error}: The algebraic derivations for each subspace can differ significantly due to subtle distinctions between PDE terms, creating a high potential for human error. For example, the spatial variable $x$ appears in both the time derivative and the diffusion term, but the corresponding coefficient matrices are different due to the subtle difference between the differential operators. Such nuances require detailed derivations for each term, increasing the likelihood of algebraic errors. Moreover, since there is an additional $\alpha$ in the diffusion term $\int_{\Omega}\nabla_{x}\delta u\cdot \alpha\nabla_{x}udx d\alpha dt$, the resulting parametric matrix is, therefore, $K_{n_\alpha n_\alpha'}^{[N_\alpha \alpha N_\alpha]} = \int_{\Omega_\alpha} \widetilde{N}_{n_\alpha}^{[\alpha]}(\alpha) \alpha \widetilde{N}_{n_\alpha'}^{[\alpha]}(\alpha) \, d\alpha$ instead of $K_{n_\alpha n_\alpha'}^{[N_\alpha  N_\alpha]} = \int_{\Omega_\alpha} \widetilde{N}_{n_\alpha}^{[\alpha]}(\alpha) \widetilde{N}_{n_\alpha'}^{[\alpha]}(\alpha) \, d\alpha$. This new weighted mass matrix is not a pre-built component in the standard FEM.

(3) \textbf{Lack of Generalizability}: The example problem illustrates a straightforward case that can be solved using TAPS. However, in more general scenarios involving additional parametric variables or new equation terms, the structure of the resulting matrix system is fundamentally altered. This alteration necessitates a complete re-derivation of the solver. Consequently, as an intrusive method, TAPS lacks the flexibility to easily adapt to different problem configurations.

(4) \textbf{Labor-Intensive Process}: The step-by-step mathematical derivation required by TAPS is not only intricate but also highly repetitive, demanding manual execution. For parametric PDEs, where the input space can include numerous material properties, forcing terms, or boundary conditions, this process is particularly inefficient. Consequently, any modification to these parameters requires the entire tedious derivation to be repeated from scratch. This manual re-derivation, coupled with the subsequent coding and verification effort, renders TAPS a labor-intensive and intrusive approach to model general parametric PDEs in CAE workflows.

Despite these challenges, we remark that the TAPS framework is inherently systematic and modular, as the entire process is defined by formal languages of mathematics and computer code. This structured nature makes TAPS an ideal candidate for automation using LLMs, which excel at both symbolic reasoning and code generation. Leveraging these capabilities can overcome the aforementioned challenges associated with the manual implementation of TAPS for various problems.

\subsection{LLM-empowered mathematical reasoning for TAPS}
While LLMs have achieved remarkable success in general language tasks, their performance deteriorates in specialized scientific domains such as data-free MOR. This deficit arises because the statistical, pattern-matching nature of LLMs is fundamentally misaligned with the rigorous derivation and numerical implementation required by the latest MOR techniques such as TAPS. Successfully applying these methods demands a precise sequence of symbolic and algebraic operations, from deriving weak forms to executing tensor contractions, which the probabilistic nature of a general-purpose LLM cannot guarantee.

Consequently, when applied to MOR solver development, LLMs often produce generic responses referencing well-established methods like Proper Orthogonal Decomposition (POD) \cite{berkooz1993proper}, or generate code and equations that are syntactically invalid or conceptually flawed. This phenomenon is known as hallucination \cite{kalai2025languagemodelshallucinate}. This capability gap renders general-purpose LLMs unreliable for direct use in sophisticated scientific workflows in CAE.

\begin{figure}[!hbt]
\centering
\includegraphics[width=0.65\linewidth]{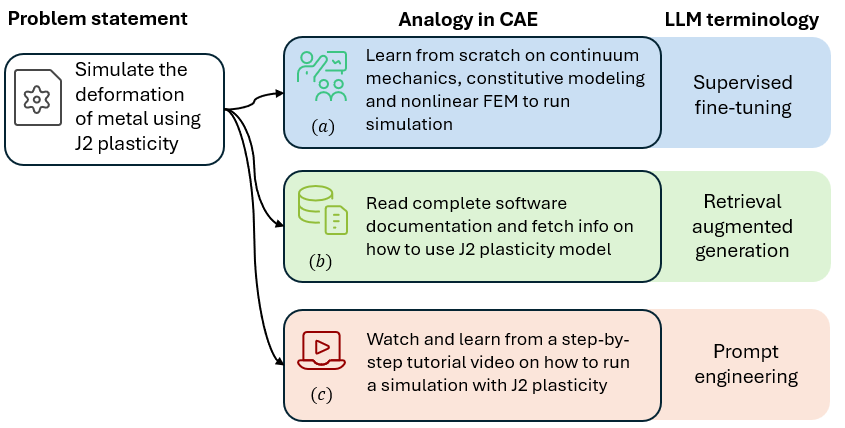}
\caption{Countering LLM hallucinations with: (a) supervised fine-tuning (SFT): fine-tune the weights of LLMs by minimizing the prediction loss on domain-specific data. An analogy in CAE is to try to learn fundamental knowledge before using the simulation software. This process is the most expensive case. (b) retrieval augmented generation (RAG): fetch relevant information from the database to assist model prediction. An analogy is to directly look into specific documentations rather than learning from scratch (c) prompt engineering: embed a similar example directly in the prompt message sent to LLMs. An analogy is that, rather than searching the relevant document, directly learn a similar CAE example step-by-step.}
\label{hallucination}
\end{figure}

Various methods exist to mitigate LLM hallucination, including supervised fine-tuning (SFT), retrieval-augmented generation (RAG), and prompt engineering, with analogies in the domain of CAE shown in Fig. \ref{hallucination}. SFT adapts a general-purpose LLM to a specific domain, but this process is often prohibitively expensive due to the need for vast amounts of high-quality training data. RAG enhances the LLM by retrieving information from a domain-specific database, though its reliability can be compromised if the retrieval process fetches irrelevant context. Prompt engineering, which involves crafting specialized prompts, is highly dependent on prompt quality, but does not require extensive domain-specific data. Given its cost-effectiveness and minimal data requirements, this paper employs a specific prompt engineering technique known as few-shot prompting to guide the LLMs \cite{wei2022chain}.

\subsubsection{Few-shot prompting}

\begin{figure}[!hbt]
\centering
\includegraphics[width=0.65\linewidth]{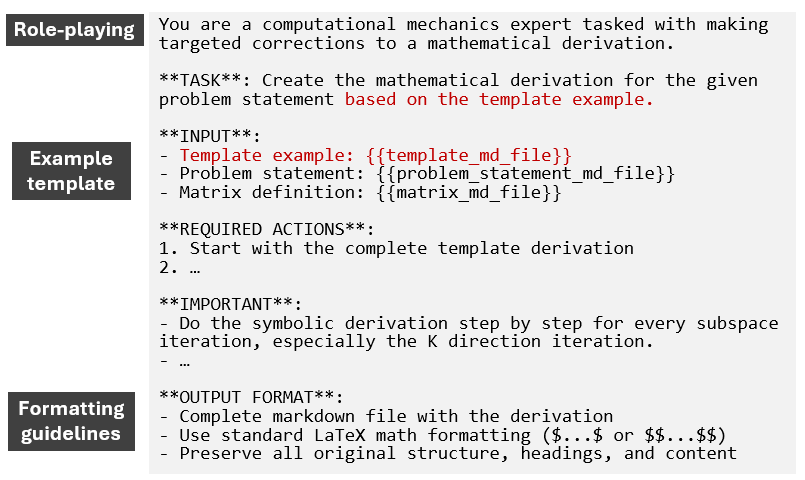}
\caption{Few-shot plotting example used for the mathematical reasoning prompt for TAPS development.}
\label{fewshot}
\end{figure}

The few-shot prompt leverages the in-context learning capabilities of modern LLMs by providing a small number of demonstration examples, or ``shots", directly within the prompt. This approach requires no updates to the LLMs' weights; instead, it guides LLMs to generate the desired output by establishing the necessary structural and syntactic patterns from the examples provided. Through this method, a general-purpose LLM can be adapted to produce highly structured domain-specific output, effectively bridging the gap between its generalized capabilities and the precise symbolic requirements within the TAPS framework. A template for this few-shot prompting approach is shown in Fig. \ref{fewshot}, where the example template serves as a demonstration of the mathematical formulation. In this paper, we used the complete mathematical derivation of the TAPS solver for Eq. \ref{pde_eqn1} as the few-shot example. The complete prompt is detailed in \ref{app:prompt}, and the Langchain library is used for fine-grained control of LLMs \cite{topsakal2023creating}.

\subsubsection{Chain-of-thought prompting}
\begin{figure}[!hbt]
\centering
\includegraphics[width=0.75\linewidth]{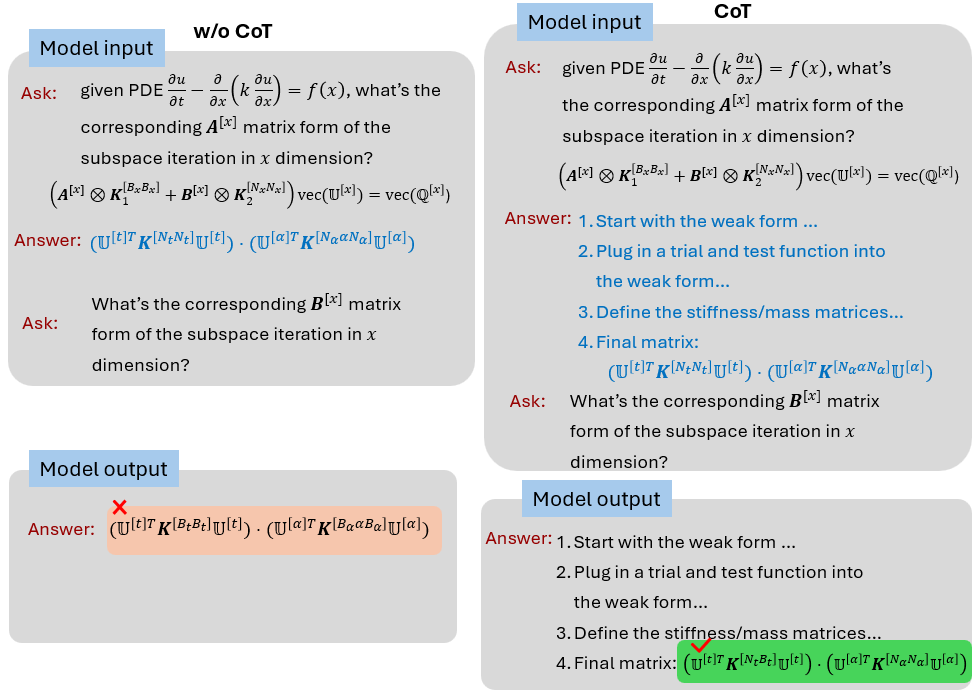}
\caption{CoT can reduce the hallucinations of LLM in deriving the matrix form of each subspace iteration by teaching LLMs step-by-step.}
\label{cot}
\end{figure}

To handle the complex tensorial symbolic derivations required by TAPS, we enhance the example template using Chain-of-Thought (CoT) prompting \cite{wei2022chain}. Unlike the standard few-shot prompt, which only provides the final answer in its examples, CoT enriches the prompt with the intermediate, step-by-step reasoning needed to arrive at the solution. This distinction is illustrated in Fig. \ref{cot}.

As can be seen, LLMs tend to hallucinate when the few-shot examples directly output the final results rather than explaining the details of how the corresponding matrix form is derived. In contrast, the CoT prompt details each step on how the weak form is derived, followed by plugging the C-HiDeNN-TD approximation of the trial and test functions into the weak form. Then the required stiffness/mass matrices are defined before the final linear system of equations is derived. By providing this explicit, step-by-step reasoning, the LLM learns to replicate the derivation process for new, unseen parametric PDEs.

\begin{figure}[!hbt]
\centering
\includegraphics[width=0.5\linewidth]{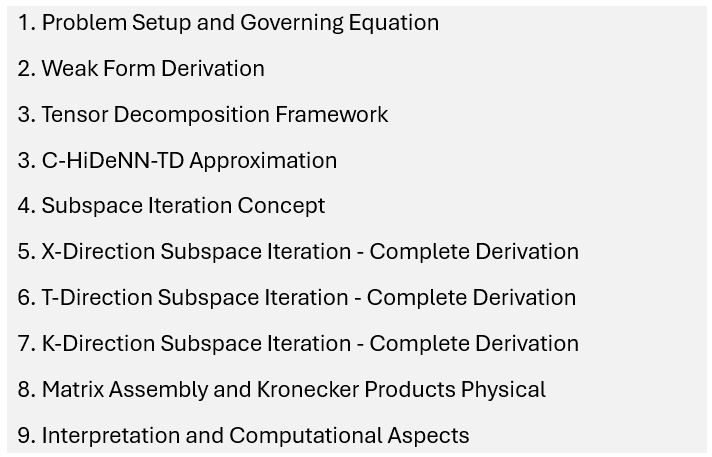}
\caption{CoT example used in the current approach: detailed step-by-step mathematical reasoning process from original S-P-T PDE to the final matrix equations.}
\label{structure}
\end{figure}

From the above reasons, the provided CoT example plays a critical role in enabling the LLM to extrapolate the TAPS derivation to more general parametric PDEs. In this paper, we prepared the mathematical derivation of TAPS for Eq. \ref{pde_eqn1} as the CoT example. The content, structured as shown in Fig. \ref{structure}, is written in Markdown format. 


\subsection{Automated agentic workflow}
 The development of CAE solvers consists of multiple procedures \cite{maric2022research}. These include mathematical derivation, high-performance computational code implementation, and verification/validation/testing, where each step requires domain expertise and can be extremely arduous. With the help of LLMs, this labor-intensive development work can be significantly alleviated. For example, there are many powerful LLM-based coding assistants that have shown exceptional performance in a vast amount of coding benchmarks \cite{jimenez2023swe, aleithan2024swe}. These coding assistants can potentially accelerate the software development process. 
 
\begin{figure}[!hbt]
\centering
\includegraphics[width=0.23\linewidth]{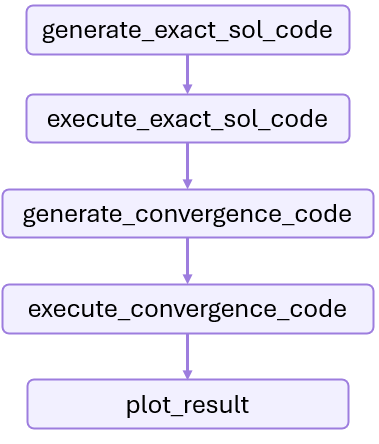}
\caption{Automated agentic workflow for verification using different MCP tools.}
\label{veri}
\end{figure}

 Once the TAPS solver code for the new parametric PDE is implemented, the verification process can be automated through MCP tools, as shown in Fig. \ref{veri}. This involves multiple stages. First, LLMs are called to generate a symbolic programming code to manufacture exact solutions to the corresponding PDE given suitable initial and boundary conditions. The LLMs then call this exact solution generator an MCP tool to generate the exact solution with its corresponding $L^2$ norm and forcing function. Subsequently, this information is inserted into a convergence study script to calculate the relative $L^2$ norm error. Finally, the LLM executes this script and automatically summarizes and plots the final results with the provided MCP tools.

\begin{figure}[!hbt]
\centering
\includegraphics[width=0.8\linewidth]{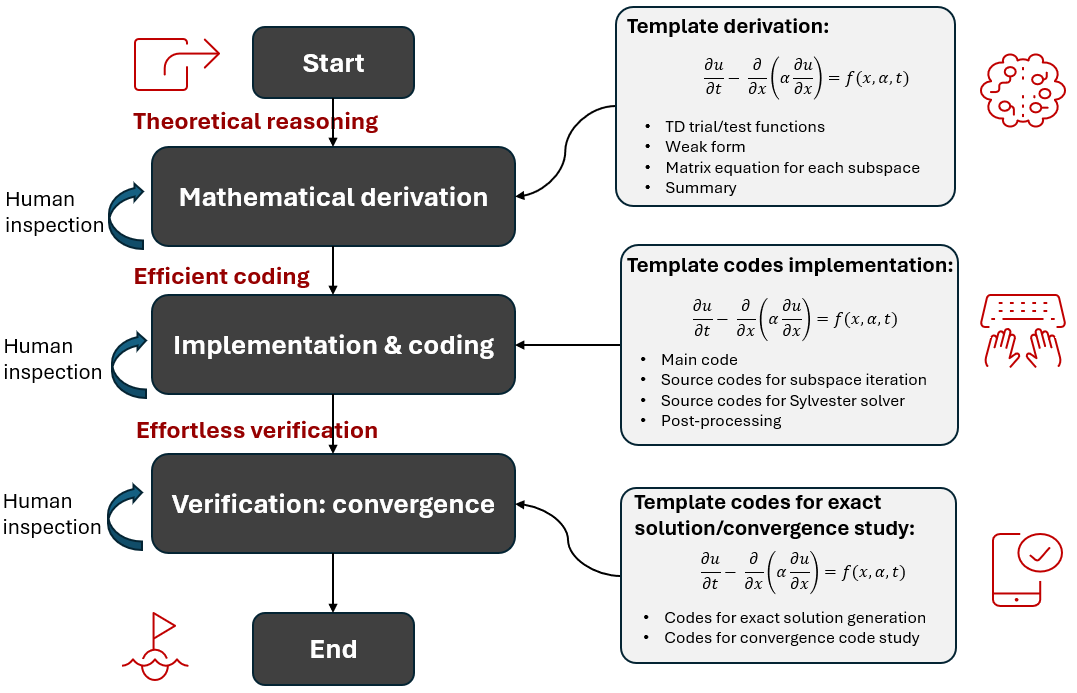}
\caption{Complete agentic workflow for data-free TAPS development empowered by LLM.}
\label{agent}
\end{figure}

 As a result, LLMs can be used to automate the whole process from theory development, coding, and verification, as shown in Fig. \ref{agent}. Although this agentic workflow can significantly accelerate the development, it should be noted that the inherently probabilistic nature of LLMs means that their outputs can be unreliable or deviate from the user's intent \cite{weisz2025examining}. Consequently, human inspection is still essential for each procedure described in Fig. \ref{agent}.

\section{Numerical results}
\label{subsec:results}
In this section, we illustrate the LLM-empowered TAPS solver for different governing equations. For all the cases investigated in this section, GPT-5 is adopted as the LLM model. The 1D S-P-T problem shown in Eq. \ref{pde_eqn1} is used as the original template problem, and the complete mathematical derivation, implementation, and convergence study codes are provided as templates in the crafted prompts. For all the numerical problems presented, mathematical derivation takes around 5 minutes; implementing numerical codes takes around 10 minutes; generating convergence study codes takes about 5 minutes. A single NVIDIA RTX A6000 GPU with 48 GB memory is used for the numerical study.

\subsection{Magnetostatics}
In this example, LLM is used to automatically derive the mathematical formulation of TAPS for the 3D magnetostatics equation. The numerical code is implemented using LLM according to the LLM-generated mathematical derivation. Finally, the convergence of the newly developed TAPS solver is investigated using the proposed automated workflow as shown in Fig. \ref{veri}. The 3D magnetostatic equation in component form is as follows:

\begin{equation}
\begin{cases}
\nabla^2 A_x(x,y,z) = -\mu_0 J_x(x,y,z) \\
\nabla^2 A_y(x,y,z) = -\mu_0 J_y(x,y,z) \\
\nabla^2 A_z(x,y,z) = -\mu_0 J_z(x,y,z)
\end{cases}
\label{eq:mag}
\end{equation}
where $\mu_0$ represents the vacuum permeability; $\bm{J} = (J_x, J_y, J_z)$ refers to the current density vector.
The solution field is approximated using C-HiDeNN-TD as follows:
\begin{subequations}
\begin{gather}
A_x^{TD}(x, y, z) = \sum_{m=1}^M A_{xx}^{(m)}(x) A_{xy}^{(m)}(y) A_{xz}^{(m)}(z) \\
A_y^{TD}(x, y, z) = \sum_{m=1}^M A_{yx}^{(m)}(x) A_{yy}^{(m)}(y) A_{yz}^{(m)}(z) \\
A_z^{TD}(x, y, z) = \sum_{m=1}^M A_{zx}^{(m)}(x) A_{zy}^{(m)}(y) A_{zz}^{(m)}(z) 
\end{gather}
\end{subequations}

Deriving the matrix form for Eq. \ref{eq:mag} using the provided simple 1D S-P-T example presents two main challenges for an LLM. First, the target equation describes a vector field, whereas the provided example is a scalar PDE. Second, the Laplacian operator is applied to a 3D spatial field, not a 1D field as in the template.

Despite these complexities, with the given prompt, GPT-5 accurately derives the TAPS formulation in a single shot. This success is attributable to two key factors: GPT-5's ability to correctly derive the weak form for PDEs in a 3D spatial domain, and its understanding of essential tensor algebra concepts, including the Einstein summation convention, contraction, vectorization, and the Kronecker product. Consequently, with a well-formulated example template for a simpler PDE, GPT-5 can extrapolate the required mathematical concepts to a more complex problem.

Once the complete and detailed mathematical formulation of the TAPS solver is derived for the new PDE, the next step is to transform the final matrix equation in each subspace iteration into the corresponding code. To this end, GPT-5 is used again to generate the TAPS solver code. It should be noted that we do not ask GPT-5 to develop the new code from scratch. Instead, we provide the complete implementation of the 1D S-P-T TAPS code for Eq. \ref{pde_eqn1} as the template. This can significantly facilitate code generation while minimizing potential errors. With the newly developed TAPS code, we investigate the convergence of the solver by leveraging a fully automated workflow in Fig. \ref{veri}. To this end, the relative $L^2$ norm error is defined as shown below:

\begin{equation}
    \epsilon_{L^2}=\frac{\|\bm{A}^{TD}(x,y,z)-\bm{A}^{\text{ex}}(x,y,z)\|_{L_2}}{\|\bm{A}^{\text{ex}}(x,y,z)\|_{L_2}}
\label{eq:l2norm_spt}
\end{equation}

\begin{figure}[!hbt]
\centering
\includegraphics[width=0.8\linewidth]{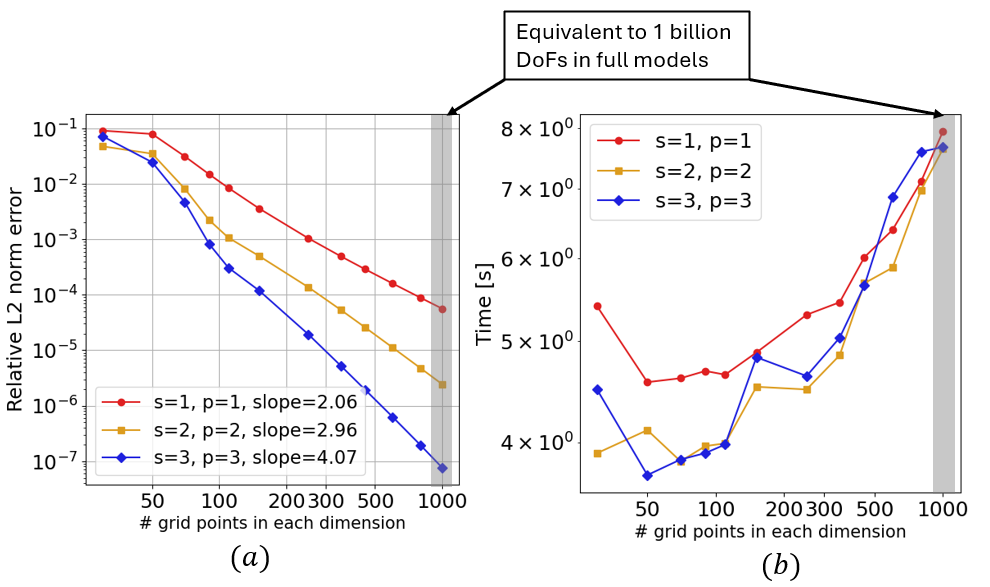}
\caption{Convergence and computational time plots for the 3D magnetostatics.}
\label{mag}
\end{figure}

The relative $L^2$ norm error and the computation time for different cases are plotted in Fig. \ref{mag}. As can be observed from the figure, the convergence rate of the developed solver depends on the C-HiDeNN hyperparameter $p$. With a higher reproducing polynomial order $p$, the accuracy of the solution can be significantly improved. For the last case used in the convergence study, we use $1,000$ grid points in each spatial direction, which is equivalent to 1 billion DoFs for a full-order model. Nevertheless, TAPS can solve this case in 8 seconds. This corroborates the significant speedup of TAPS as an MOR solver compared to standard full-order models such as FEM.

\subsection{Linear elasticity}
In this example, we use LLM to perform mathematical derivation, numerical implementation, and convergence study of the 3D linear elasticity problem. The governing equation is defined as:
\begin{equation}
\begin{cases}
\mu \nabla^2 u(x,y,z) + (\lambda + \mu) \frac{\partial}{\partial x}\left[\nabla \cdot \mathbf{u}(x,y,z)\right] + f_x(x,y,z) = 0 \\

\mu \nabla^2 v(x,y,z) + (\lambda + \mu) \frac{\partial}{\partial y}\left[\nabla \cdot \mathbf{u}(x,y,z)\right] + f_y(x,y,z) = 0 \\

\mu \nabla^2 w(x,y,z) + (\lambda + \mu) \frac{\partial}{\partial z}\left[\nabla \cdot \mathbf{u}(x,y,z)\right] + f_z(x,y,z) = 0
\end{cases}
\end{equation}

where $(\lambda, \mu)$ represent Lamé parameters. The solution field is approximated with C-HiDeNN-TD:
\begin{subequations}
\begin{gather}
u^{TD}(x, y, z) = \sum_{m=1}^M u_{x}^{(m)}(x) u_{y}^{(m)}(y) u_{z}^{(m)}(z) \\
v^{TD}(x, y, z) = \sum_{m=1}^M v_{x}^{(m)}(x) v_{y}^{(m)}(y) v_{z}^{(m)}(z) \\
w^{TD}(x, y, z) = \sum_{m=1}^M w_{x}^{(m)}(x) w_{y}^{(m)}(y) w_{z}^{(m)}(z)
\end{gather}
\end{subequations}

As can be readily seen, the main challenge in the mathematical derivation of LLMs for this case is the dilational term $(\lambda + \mu) \nabla\left[\nabla \cdot \mathbf{u}(x,y,z)\right]$, as it will result in coupled terms from different dimensions in the C-HiDeNN-TD approximation. However, the techniques for handling terms like this are not present in the template file of Eq. \ref{pde_eqn1}. Despite this, GPT-5 still manages to derive the final matrix form correctly. For example, the derived matrix form for the subspace iteration in $u_y$ dimension is:

\begin{equation}
\left[ \mu \cdot \bm{C}_{uu,lap}^{[y,x]} \otimes \bm{M}^{[y]}  +  \mu \cdot \bm{C}_{uu,lap}^{[y,y]} \otimes \bm{K}^{[y]} + (\mu \cdot \bm{C}_{uu,lap}^{[y,z]} + (\lambda + \mu) \cdot \bm{C}_{uu,div}^{[y]})\otimes \bm{M}^{[y]}  \right] \text{vec}(\mathbb{U}^{[y]}) = \text{vec}(\mathbb{Q}_{u_y}^{[y]})
\end{equation}

where the coefficient matrices are defined as:
\begin{subequations}
\begin{gather}
\bm{C}_{uu,lap}^{[y,x]} = \left(\mathbb{U}^{[x]T} \bm{K}^{[x]} \mathbb{U}^{[x]}\right) \odot \left(\mathbb{U}^{[z]T} \bm{M}^{[z]} \mathbb{U}^{[z]}\right)  \\
\bm{C}_{uu,lap}^{[y,y]}= \left(\mathbb{U}^{[x]T} \bm{M}^{[x]} \mathbb{U}^{[x]}\right) \odot \left(\mathbb{U}^{[z]T} \bm{M}^{[z]} \mathbb{U}^{[z]}\right)\\
\bm{C}_{uu,lap}^{[y,z]}= \left(\mathbb{U}^{[x]T} \bm{M}^{[x]} \mathbb{U}^{[x]}\right) \odot \left(\mathbb{U}^{[z]T} \bm{K}^{[z]} \mathbb{U}^{[z]}\right)\\
\bm{C}_{uu,div}^{[y]}= \left(\mathbb{U}^{[x]T} \bm{K}^{[x]} \mathbb{U}^{[x]}\right) \odot \left(\mathbb{U}^{[z]T} \bm{M}^{[z]} \mathbb{U}^{[z]}\right)
\end{gather}
\end{subequations}

The forcing term is composed of 3 parts $\mathbb{Q}^{[y]} = \mathbb{Q}^{[y]} + \mathbb{Q}^{[y,uv]} + \mathbb{Q}^{[y,uw]}$ where $\mathbb{Q}^{[y]}$ is the integration of the multiplication of the C-HiDeNN function and the forcing term, and the other two terms are defined as:
\begin{subequations}
\begin{gather}
\mathbb{Q}^{[y,uv]} = -(\lambda + \mu)  \bm{K}^{[y,mix]T} \mathbb{V}^{[y]}  \bm{C}_{uv,div}^{[y]}\\
\mathbb{Q}^{[y,uw]} = -(\lambda + \mu) \bm{M}^{[y]}  \mathbb{W}^{[y]}  \bm{C}_{uw,div}^{[y]}
\end{gather}
\end{subequations}

with two additional new coefficient matrices defined as:
\begin{subequations}
\begin{gather}
\bm{C}_{uv,div}^{[y]}=  \left(\mathbb{U}^{[x]T} \bm{K}^{[x,mix]} \mathbb{V}^{[x]}\right) \odot \left(\mathbb{U}^{[z]T} \bm{M}^{[z]} \mathbb{V}^{[z]}\right) \\
\bm{C}_{uw,div}^{[y]}= \left(\mathbb{U}^{[x]T} \bm{K}^{[x,mix]} \mathbb{W}^{[x]}\right) \odot \left(\mathbb{U}^{[z]T} \bm{K}^{[z,mix]T} \mathbb{W}^{[z]}\right) 
\end{gather}
\end{subequations}

As can be seen from the above equation, a new type of assembled matrix has been introduced.
\begin{equation}
\bm{K}^{[d,mix]} = \int_{\Omega_d} \widetilde{\bm{B}}^{[d]T}(x_d) \widetilde{\bm{N}}^{[d]}(x_d) \, dx_d
\end{equation}
$\bm{K}^{[d,mix]}$ is resulted from dilational term in the weak form. Although it is not shown in the previous template derivation, GPT-5 manages to properly define this new matrix and derive the correct corresponding matrix form. This again manifests the mathematical reasoning capability and flexibility of using LLMs for MOR solver development. LLMs can extend the provided simple example to more complex cases based on their internal knowledge base.

\begin{figure}[!hbt]
\centering
\includegraphics[width=0.8\linewidth]{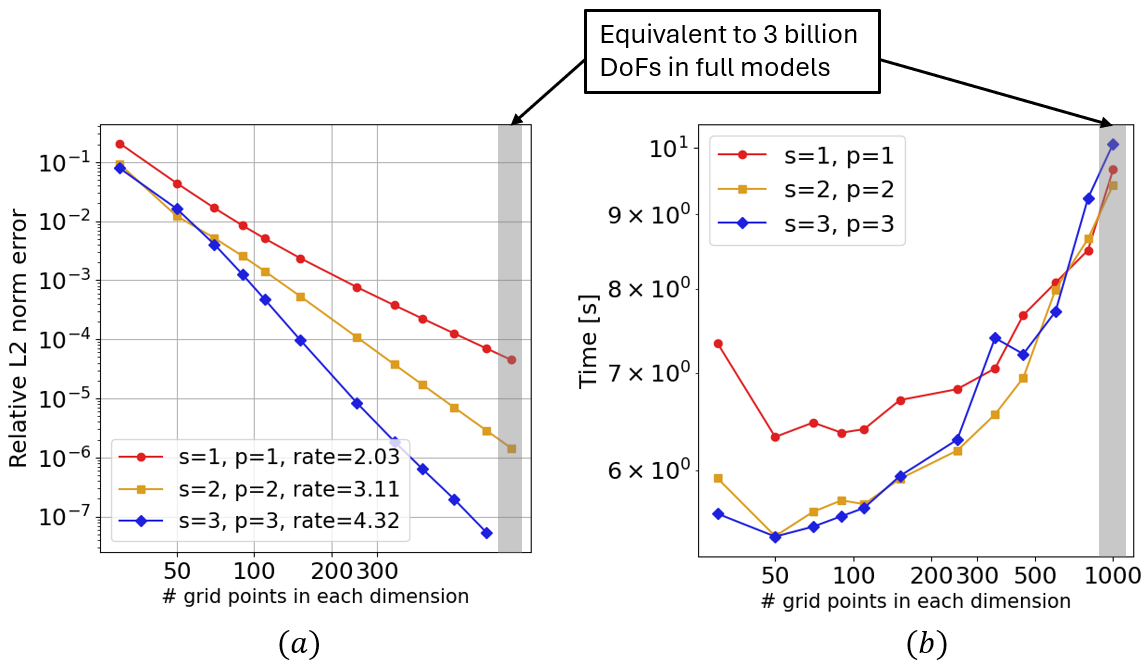}
\caption{Convergence and computational time plots for the 3D elasticity.}
\label{elastic}
\end{figure}

With the newly developed TAPS code for elasticity, the convergence and speed of the solver are shown in Fig. \ref{elastic}. As expected, the convergence rate is controlled by the reproducing polynomial order $p$. When $1,000$ grid points are used for each dimension (equivalent to 3 billion DoFs for a full-order model), the TAPS solver takes only about 10 seconds to complete the simulation.

\subsection{Nonlinear PDE}
In this example, we use LLMs to extend the original TAPS code for linear problems to a nonlinear case. The 3D transient diffusion-reaction equation is shown in Eq. \ref{eq:nonlinear}.

\begin{equation}\frac{\partial u(x,y,z)}{\partial t} - \bm{\nabla} \cdot \left(\alpha \bm{\nabla}{u}\right) + u^2 = f
\label{eq:nonlinear}
\end{equation}

Instead of solving a standard space-time problem, we treat diffusivity $\alpha$ as the parametric variable, and the S-P-T approximation to the parametric solution is: 
\begin{equation}
u^{TD}(x, y, z, \alpha, t) = \sum_{m=1}^M u_x^{(m)}(x) u_y^{(m)}(y) u_z^{(m)}(z) u_\alpha^{(m)}(\alpha) u_t^{(m)}(t)
\end{equation}

\begin{figure}[!hbt]
\centering
\includegraphics[width=0.8\linewidth]{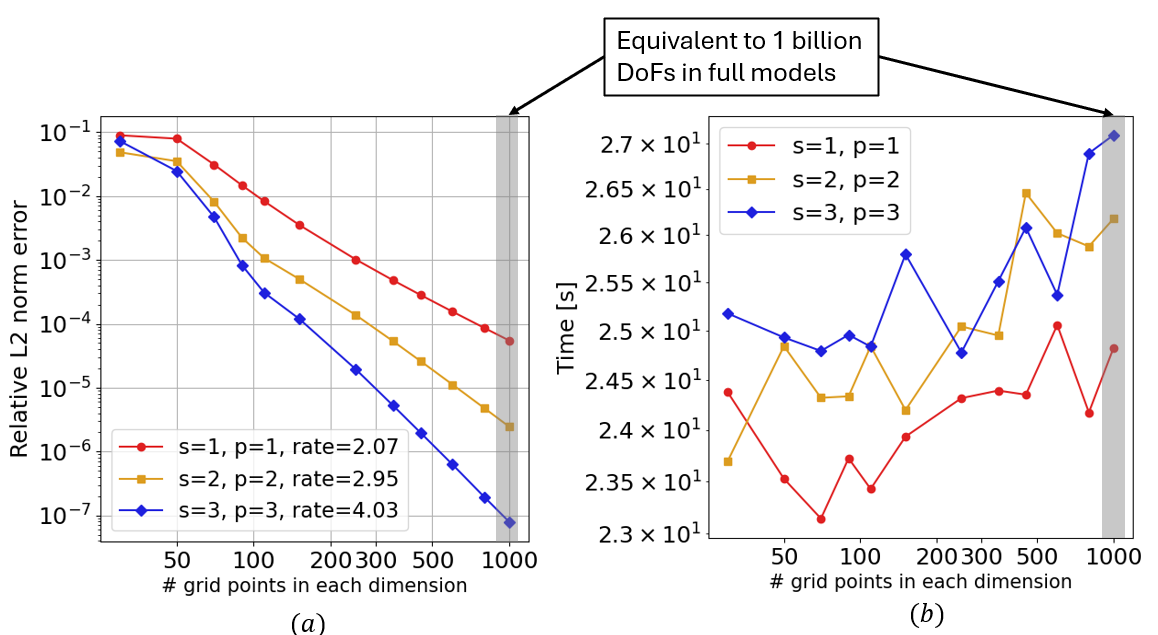}
\caption{Convergence plot for the 3D nonlinear PDE problem.}
\label{nonlinear}
\end{figure}

The main challenge for LLMs in this problem is the treatment of the nonlinear reaction term $u^2$. Even though the TAPS solver has not been used to solve nonlinear PDEs before, when Eq. \ref{nonlinear} is used as input, GPT-5 recognizes the nonlinearity of the problem and recommends using the fixed-point iteration method to address nonlinearity. This is because GPT-5 has a good understanding of general solution schemes to nonlinear PDEs and can actively adapt them to the current problem. As suggested by GPT-5, the linearized equation in each iteration can be written as:
\begin{equation}
\frac{\partial u^{(n+1)}}{\partial t} - \nabla_{x,y,z}\cdot\left(k \nabla_{x,y,z} u^{(n+1)}\right) + u^{(n)} u^{(n+1)} = f
\end{equation}
where $n$ is the index for the nonlinear iteration.

When GPT-5 was initially used for the mathematical derivation, it struggled to handle the integration of the linearized reaction term efficiently, as this form is atypical in the standard finite element:

\begin{equation}
\int_{\Omega} \delta u^{(i+1)} u^{(i)} u^{(i+1)} \, dxdydzd\alpha dt
\end{equation}

As a result, we manually defined a new type of matrix: a weighted mass matrix as input to the \textit{matrix definition} part of the input prompt, as shown in Fig. \ref{fewshot}.

\begin{equation}
\bm{M}^{[d](i)} = \int_{\Omega_{d}} \widetilde{\bm{N}}^{[d]T}(x_d)\, u_{d}^{(i)}(x_d)\, \widetilde{\bm{N}}^{[d]}(x_d)\, dx_d
\end{equation}
where $d$ refers to the dimension index.

With this new matrix definition, the GPT-5 model successfully derived the correct final matrix form for each subspace iteration of the TAPS solver. For example, the final matrix form of the subspace iteration in the $z$ direction is derived as:

\begin{equation}
\Big[ \bm{C}_{x,z}^{[z]}  \otimes  \bm{K}^{[z]}
\;+\; (\bm{C}_{x,x}^{[z]} + \bm{C}_{x,y}^{[z]} + \bm{C}_{t}^{[z]})\otimes  \bm{M}^{[z]} \Big]\, \text{vec}(\mathbb{U}^{[z]})
\;+\; \left[ \sum_{p=1}^M \left( \bm{\Gamma}^{[z](p)} \otimes\bm{M}^{[z](p)}  \right) \right] \text{vec}(\mathbb{U}^{[z]})
= \text{vec}(\mathbb{Q}^{[z]})
\end{equation}

where the coefficient matrices are defined as:
\begin{subequations}
    
\begin{gather}
\bm{C}_{x,z}^{[z]} =
\left(\mathbb{U}^{[x]T} \bm{M}^{[x]} \mathbb{U}^{[x]}\right) \odot
\left(\mathbb{U}^{[y]T} \bm{M}^{[y]} \mathbb{U}^{[y]}\right) \odot
\left(\mathbb{U}^{[\alpha]T} \bm{K}^{[N_\alpha \alpha N_\alpha]}  \mathbb{U}^{[\alpha]}\right)\odot
\left(\mathbb{U}^{[t]T} \bm{K}^{[N_t N_t]} \mathbb{U}^{[t]}\right)\\
\bm{C}_{x,x}^{[z]}=  \left(\mathbb{U}^{[x]T} \bm{K}^{[x]} \mathbb{U}^{[x]}\right) \odot \left(\mathbb{U}^{[y]T} \bm{M}^{[y]} \mathbb{U}^{[y]}\right) \odot \left(\mathbb{U}^{[\alpha]T} \bm{K}^{[N_\alpha \alpha N_\alpha]} \mathbb{U}^{[\alpha]}\right) \odot \left(\mathbb{U}^{[t]T} \bm{K}^{[N_t N_t]} \mathbb{U}^{[t]}\right)
 \\
\bm{C}_{x,y}^{[z]} =  \left(\mathbb{U}^{[x]T} \bm{M}^{[x]} \mathbb{U}^{[x]}\right) \odot \left(\mathbb{U}^{[y]T} \bm{K}^{[y]} \mathbb{U}^{[y]}\right) \odot \left(\mathbb{U}^{[\alpha]T} \bm{K}^{[N_\alpha \alpha N_\alpha]} \mathbb{U}^{[\alpha]}\right) \odot \left(\mathbb{U}^{[t]T} \bm{K}^{[N_t N_t]} \mathbb{U}^{[t]}\right)
 \\
\bm{C}_{t}^{[z]}= \left(\mathbb{U}^{[x]T} \bm{M}^{[x]} \mathbb{U}^{[x]}\right) \odot \left(\mathbb{U}^{[y]T} \bm{M}^{[y]} \mathbb{U}^{[y]}\right) \odot \left(\mathbb{U}^{[\alpha]T} \bm{K}^{[N_\alpha N_\alpha]} \mathbb{U}^{[\alpha]}\right) \odot \left(\mathbb{U}^{[t]T} \bm{K}^{[N_t B_t]} \mathbb{U}^{[t]}\right)
\end{gather}
\end{subequations}

This example shows that human intervention is still required for some unseen, complicated mathematical operations. Similarly, the convergence and computation cost plots are shown in Fig. \ref{nonlinear}. The convergence rates for different C-HiDeNN hyperparameters follow the same pattern. When $1,000$ grid points are used for each dimension, the total equivalent DoFs for a full-order model are 1 billion. However, the nonlinear TAPS solver only takes around $27$ seconds, which is much faster than the corresponding full-order models such as FEM.

\subsection{Transient heat transfer with spatially variant diffusivity}
\begin{figure}[!hbt]
\centering
\includegraphics[width=0.3\linewidth]{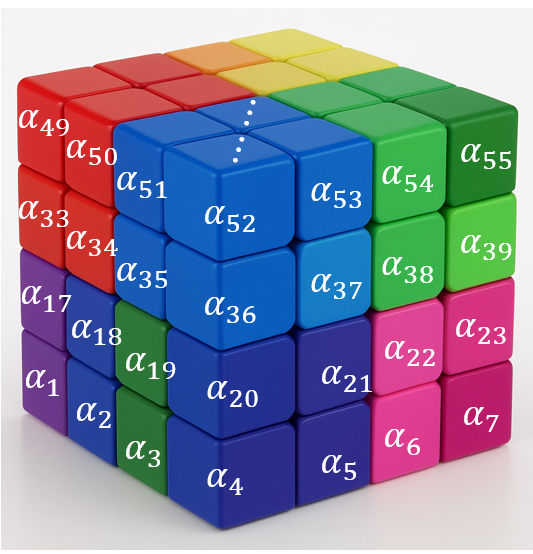}
\caption{3D parametric material design problem.}
\label{dd3d}
\end{figure}

In this example, our aim is to leverage LLMs to develop the TAPS solver for the following problem:
\begin{equation}
\frac{\partial u\left[\bm{x}, \alpha(\bm{x}), t\right]}{\partial t} - \mathbf{\nabla} \cdot \Big[\alpha(\bm{x}) \nabla{u}\left[\bm{x}, \alpha(\bm{x}), t\right]\Big] = f
\label{eqdd}
\end{equation}
where the diffusivity field $\alpha(\bm{x})$ can vary at different locations. As shown in Fig. \ref{dd3d}, the whole domain is decomposed into $D_x \times D_y \times D_z$ subdomains where the diffusivity is homogeneous in each subdomain but can arbitrarily change within a predefined range $[\alpha_{min}, \alpha_{max}]$. The parametric input variable $\bm{\alpha} = (\alpha_1, \alpha_2, \cdots, \alpha_i, \cdots,\alpha_{D_xD_yD_z})$ where $\alpha_i$ stands for the diffusivity for the $i$-th subdomain. Therefore, the solution to this problem is a S-P-T field and can be approximated using TD as follows:

\begin{equation}u^{TD}(x,y,z,\alpha_1,\alpha_2, \cdots,\alpha_{D_xD_yD_z},t)=\sum_{m=1}^Mu_x^{(m)}(x)u_y^{(m)}(y)u_z^{(m)}(z)\left[\prod_{i=1}^{D_xD_yD_z}u_{\alpha_i}^{(m)}(\alpha_i)\right]u_t^{(m)}(t) \label{domain}\end{equation}

As can be seen from Eq. \ref{domain}, this is an inherently high-dimensional problem due to the large number of input parametric variables. When a fine mesh is used for the discretization, standard data-driven methods can be extremely expensive due to the sheer amount of offline data required to be generated in the large sampling space. As an alternative, TAPS does not require any offline data but directly solves the governing parametric PDE in Eq. \ref{eqdd}. However, deriving the corresponding matrix equations can be quite challenging and cumbersome due to the complex algebra involved and the flexible nature of the problem statement, as different numbers of subdomains can be used.

GPT-5 is used to derive the matrix form of each subspace iteration. In the \textit{problem statement} part of the prompt, as shown in Fig. \ref{fewshot}, apart from the governing equation, we also define the parametric heterogeneous diffusivity field as:
\begin{equation}
k(x,y,z)=\sum_{r=1}^{D_xD_yD_z} k_r\, I_r(x,y,z),
\qquad
I_r(x,y,z)=I_\alpha(x)\,I_\beta(y)\,I_\gamma(z),
\end{equation}
where $I_\alpha(x_d)$ represents the indicator function whose value is equal to 1 within the corresponding subdomain and zero elsewhere. GPT-5 successfully derives all the detailed equations from the weak form to the final matrix forms in each subspace iteration by a single shot. For example, the final matrix form for the subspace iteration in the $\alpha_q$ direction is:
\begin{equation}
\left(
\sum_{\substack{r=1\\ r\neq q}}^{D_xD_yD_z}  \bm{B}_{r\neq q}^{[k_q]}  \otimes\bm{K}^{[N_{\alpha_q} N_{\alpha_q}]}
\;+\;
\bm{B}_{r=q}^{[k_q]} \otimes \bm{K}^{[N_{\alpha_q} \alpha_q N_{\alpha_q}]} 
\;+\;
\bm{B} \otimes \bm{K}^{[N_{\alpha_q} N_{\alpha_q}]} 
\right)
\text{vec}(\mathbb{U}^{[\alpha_q]})
=
\text{vec}(\mathbb{Q}^{[\alpha_q]})
\end{equation}
where the coefficient matrices are defined as follows:
\begin{subequations}
\begin{gather}
\bm{B} = \bm{X}^{(0)} \odot \bm{Y}^{(0)} \odot \bm{Z}^{(0)} \odot \big(\bigodot_{i\neq q} \bm{G}_i^{(0)}\big) \odot \bm{T}^{(1)}\\
\bm{B}_{r=q}^{[k_q]} = \bm{T}^{(0)} \odot \bm{S}_q \odot \big(\bigodot_{i\neq q} \bm{G}_i^{(0)}\big)\\
\bm{B}_{r\neq q}^{[k_q]} = \bm{T}^{(0)} \odot \bm{S}_r \odot \bm{G}_r^{(1)} \odot \big(\bigodot_{i\neq q,r} \bm{G}_i^{(0)}\big)
\end{gather}
\label{dd_matrix}
\end{subequations}

The detailed matrices involved are defined as follows:
\begin{subequations}
    \begin{gather}
\bm{G}_i^{(0)} = \mathbb{U}^{[\alpha_i]T} \bm{K}^{[N_{\alpha_i}N_{\alpha_i}]} \mathbb{U}^{[\alpha_i]}\\
\bm{G}_i^{(1)} = \mathbb{U}^{[\alpha_i]T} \bm{K}^{[N_{\alpha_i} \alpha_i N_{\alpha_i}]} \mathbb{U}^{[\alpha_i]}\\
\bm{S}_r \;=\; \bm{X}_\alpha^{(B)} \odot \bm{Y}_\beta^{(N)} \odot \bm{Z}_\gamma^{(N)}
\;+\; \bm{X}_\alpha^{(N)} \odot \bm{Y}_\beta^{(B)} \odot \bm{Z}_\gamma^{(N)}
\;+\; \bm{X}_\alpha^{(N)} \odot \bm{Y}_\beta^{(N)} \odot \bm{Z}_\gamma^{(B)}
    \end{gather}
\end{subequations}

Temporal:
\begin{subequations}
    \begin{gather}
\bm{T}^{(0)} = \mathbb{U}^{[t]T} \bm{K}^{[N_tN_t]} \mathbb{U}^{[t]}\\
\bm{T}^{(1)} = \mathbb{U}^{[t]T} \bm{K}^{[N_tB_t]} \mathbb{U}^{[t]}
    \end{gather}
\end{subequations}

Unweighted spatial masses:
\begin{subequations}
    \begin{gather}
\bm{X}^{(0)} = \mathbb{U}^{[x]T} \bm{M}^{[x]} \mathbb{U}^{[x]}\\
\bm{Y}^{(0)} = \mathbb{U}^{[y]T} \bm{M}^{[y]} \mathbb{U}^{[y]}\\
\bm{Z}^{(0)} = \mathbb{U}^{[z]T} \bm{M}^{[z]} \mathbb{U}^{[z]}
    \end{gather}
\end{subequations}

Indicator-weighted spatial contractions:
\begin{subequations}
    \begin{gather}
\bm{X}_\alpha^{(B)} = \mathbb{U}^{[x]T} \bm{K}_x^{(\alpha)} \mathbb{U}^{[x]}, X_\alpha^{(N)} = \mathbb{U}^{[x]T} \bm{M}_x^{(\alpha)} \mathbb{U}^{[x]}\\
\bm{Y}_\beta^{(B)} = \mathbb{U}^{[y]T} \bm{K}_y^{(\beta)} \mathbb{U}^{[y]}, Y_\beta^{(N)} = \mathbb{U}^{[y]T} \bm{M}_{y}^{(\beta)} \mathbb{U}^{[y]}\\
\bm{Z}_\gamma^{(B)} = \mathbb{U}^{[z]T} \bm{K}_z^{(\gamma)} \mathbb{U}^{[z]}, Z_\gamma^{(N)} = \mathbb{U}^{[z]T} \bm{M}_z^{(\gamma)} \mathbb{U}^{[z]}
    \end{gather}
\end{subequations}

where the indicator-weighted spatial sparse matrices are defined as:
\begin{subequations}
    \begin{gather}
\bm{M}^{(\alpha)}_{x} = \int_{\Omega_{x}} \widetilde{\bm{N}}^{[x]T}(x)\, I_{\alpha}(x)\, \widetilde{\bm{N}}^{[x]}(x)\, dx\\
\bm{K}^{(\alpha)}_{x} = \int_{\Omega_{x}} \widetilde{\bm{B}}^{[x]T}(x)\, I_{\alpha}(x)\, \widetilde{\bm{B}}^{[x]}(x)\, dx
    \end{gather}
\end{subequations}

We first verify the developed TAPS solver using the following forcing function. 
\begin{equation}
    f(x,z)=e^{-\Big(\frac{x^2}{r^2}+\frac{z^2}{r^2}\Big)}
\end{equation}

\begin{figure}[!hbt]
\centering
\includegraphics[width=0.95\linewidth]{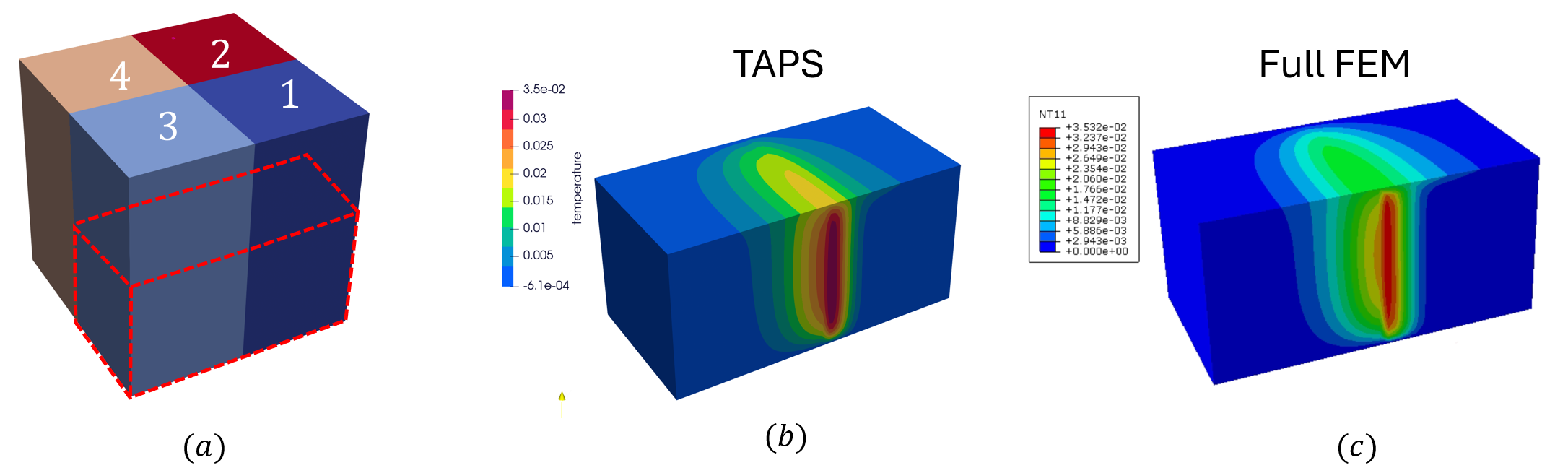}
\caption{Comparison of TAPS and full FEM prediction.}
\label{abq}
\end{figure}

Since there is no analytical solution, the results of the TAPS solver are compared with the full-order FEM simulation using Abaqus. In this comparison, the original domain is composed of $1\times2\times 2$ subdomains. The whole spatial domain is discretized with $80\times80\times80$ elements. The time $t$ and each diffusivity parameter $\alpha_p$ are also discretized using 80 elements in TAPS. In terms of computational time, Abaqus takes around $1,860$ seconds on the Intel Core i9-14900KF (3.20 GHz) CPU to predict the space-time solution for a specific set of diffusivity parameters, while TAPS takes only 601 seconds for the complete parametric simulation (any combinations of diffusivity parameters in the given range). The prediction of TAPS and the Abaqus result for the final time step are shown in Fig. \ref{abq}. As can be seen from the figure, the TAPS solution agrees well with Abaqus.

\begin{figure}[!hbt]
\centering
\includegraphics[width=0.95\linewidth]{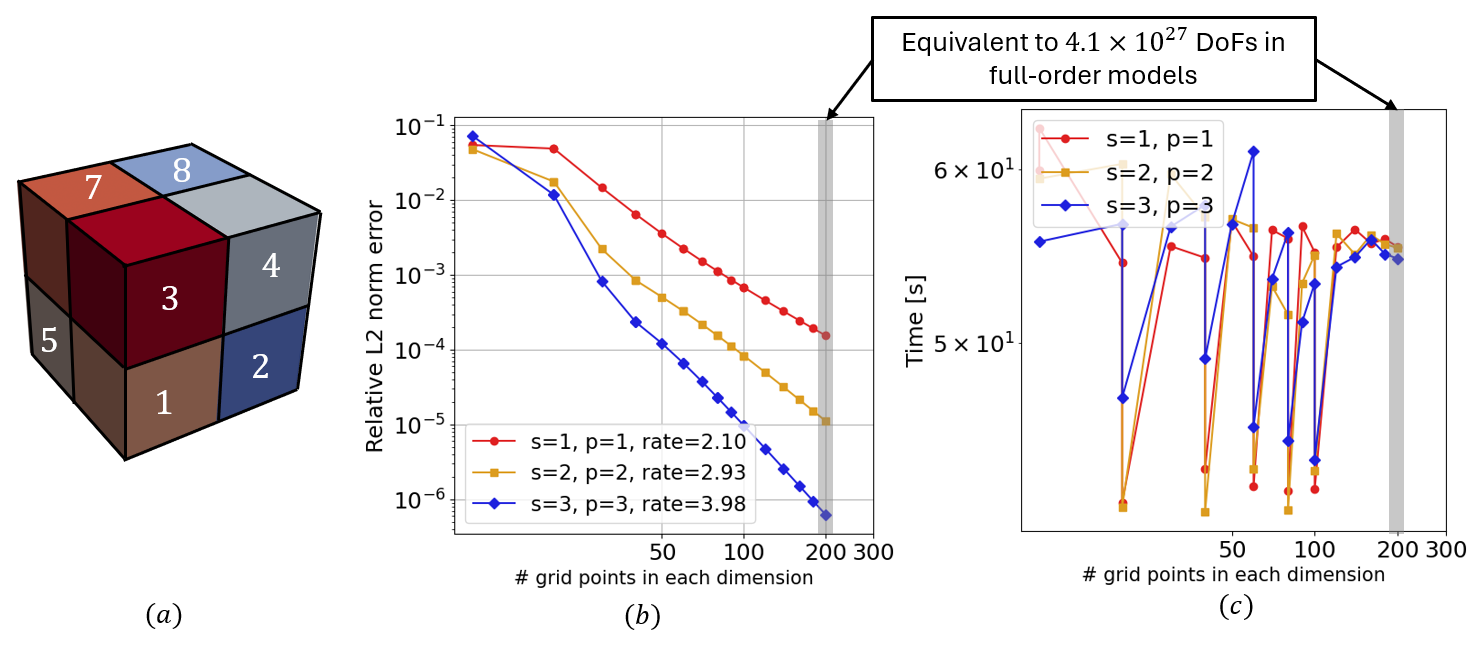}
\caption{Convergence and solution time for the case with $2\times2\times2$ subdomains.}
\label{dd3_1}
\end{figure}

To rigorously examine the accuracy of the TAPS solver, we use the automated agentic workflow for verification as shown in Fig. \ref{veri} and investigate the convergence of the developed TAPS solver. We first adopt a case with $2\times2\times2$ subdomains. The corresponding convergence of relative $L_2$ error and computation time plots are shown in Fig. \ref{dd3_1}. As expected, the parametric TAPS solver has a convergence rate of $p+1$ for different cases.

Furthermore, we also study the convergence of the TAPS solver for the case with $4\times4\times4$ subdomains. From Eq. \ref{domain}, the total number of input variables equals 68, which makes the full-order model extremely high-dimensional. However, the TAPS solver not only has the expected convergence rate, but is also significantly faster. When 200 grid points are used in all dimensions (the equivalent full-order model has $3\times10^{156}$ DoFs), the solution time is around $1,100$ seconds. This certifies the efficiency, accuracy, and flexibility of the LLM-empowered TAPS solver for large-scale, high-dimensional parametric problems.

\begin{figure}[!hbt]
\centering
\includegraphics[width=0.95\linewidth]{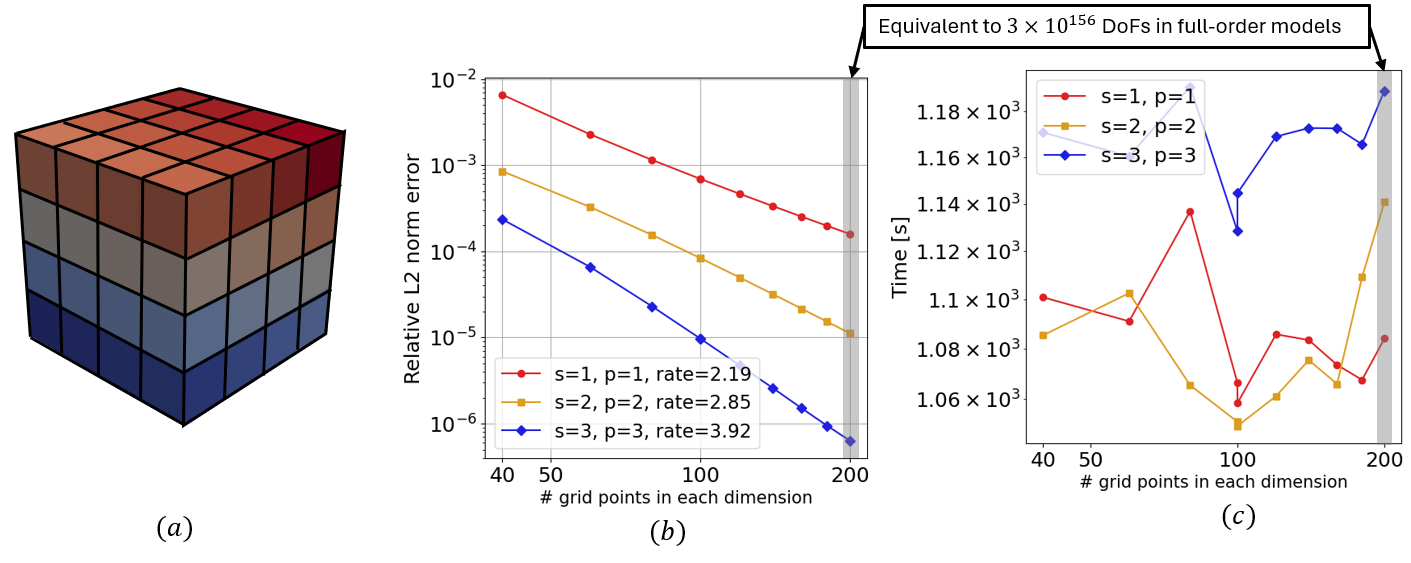}
\caption{Convergence and solution time for the case with  $4\times4\times4$ subdomains.}
\label{dd3_2}
\end{figure}

\section{Discussion}
\label{sec:discussion}
In the previous sections, we have shown the advantages of the LLM-empowered TAPS solver compared to standard numerical algorithms such as FEM in terms of computational cost. This new approach also has significant benefits in terms of RAM efficiency and disk storage requirements since the reduced-order model has fewer unknowns compared to the full-order model. Moreover, we also demonstrated that LLMs can be used to streamline the development of the TAPS solver for different complex parametric PDEs from a simple 1D S-P-T problem. This remarkable generalization capability stems from the LLM's vast pre-trained knowledge of sophisticated numerical algorithms, allowing it to synthesize novel, complex solvers from simple examples.

Compared to standard data-driven surrogates, our new approach does not require offline training data. Consequently, it has superior efficiency for large-scale problems where each full-order simulation data point can be costly to generate for data-driven methods. Moreover, TAPS is built on a weak form and has a controllable convergence rate according to the C-HiDeNN hyperparameters. Therefore, it can achieve orders of magnitude better accuracy compared to data-driven methods, which are probabilistic in nature.

In all the numerical examples, we used TAPS as the data-free MOR solver for ultra-fast, large-scale simulation of parametric PDEs. Nevertheless, note that the proposed framework is versatile and applicable to other intrusive MOR methods, such as proper orthogonal decomposition (POD), provided that a well-formatted example template derivation file and implementation code are included in the prompt for mathematical derivation.

\section{Conclusion}
\label{sec:conclusion}
In this paper, we propose a new LLM-empowered CAE agent for data-free model order reduction. This new paradigm leverages the power of LLMs and the efficiency of a data-free MOR method: TAPS. LLMs enable an automated workflow from the mathematical derivation, implementation, and verification of the TAPS solver for various kinds of parametric PDEs with minimal human intervention. This new framework can largely alleviate the challenges from the implementation of an intrusive MOR solver to different high-dimensional parametric problems, while ensuring the speed and accuracy of the developed solver. Therefore, it stands out as a promising tool for future CAE agents for ultra-fast, large-scale parametric problems that are common in engineering applications.

In the future, a mechanistic language model (MLM) can be developed by improving the general LLM with domain knowledge in the field of computational science and engineering. This can be done by fine-tuning the pre-trained language model on a curated, high-quality dataset that encompasses continuum mechanics, numerical analysis, tensor algebra, constitutive modeling, high-performance computing, and many others. Moreover, rather than using a standard gigantic transformer architecture as in general-purpose LLMs, the MLM should be of relatively smaller size since it has a relatively narrow focus on the area of CAE. With a much smaller model size, MLM will greatly improve reasoning speed, enabling near real-time mathematical derivation for time-sensitive tasks. Moreover, it can significantly reduce hardware requirements, making it more economical and applicable to typical computing resources \cite{belcak2025small}.

\section*{CRediT authorship contribution statement}
\textbf{Jiachen Guo}: Writing – original draft, Conceptualization, Methodology, \textbf{Chanwook Park}: Writing – review and editing, Data visualization. \textbf{Dong Qian}: Writing – review and editing, Project administration, Methodology. \textbf{TJR Hughes}: Writing – review and editing, Conceptualization. \textbf{Wing Kam Liu}: Writing – review and editing, Project administration, Conceptualization, Methodology.
\appendix

\section{CAE agent via model context protocol}
\label{Appendix:mcp}

The Model Context Protocol (MCP) \cite{anthropic2023mcp} is a recently proposed standard that enables LLMs to interact consistently with external tools and resources through structured messages. In contrast to simple prompt-based augmentation, MCP provides a schema-driven communication layer that ensures reproducibility, extensibility, and interoperability across diverse domains. By exposing software functionalities as MCP tools, developers can allow LLMs to dynamically discover, invoke, and combine these tools during reasoning or workflow execution. In the context of CAE, MCP makes it possible to encapsulate functions such as geometry import, mesh generation, solver execution, and surrogate modeling within a unified framework that LLMs can orchestrate autonomously.

At a practical level, the configuration of a public MCP server allows a desktop LLM agent to access specialized tools. Three common cases illustrate this process:

\begin{itemize}   

\item Claude Desktop: Anthropic’s Claude client natively supports MCP, so a server can be registered by adding the server endpoint to the configuration file. Once linked, Claude can automatically query the available tools and expose them in its interface.


\item Ollama: Ollama, when extended with an MCP bridge, can connect to local or remote MCP servers by launching with the proper --mcp-server flag. This enables LLMs running locally to call MCP tools for simulation, data processing, or visualization without relying on cloud resources.

\end{itemize}

Through these configurations, MCP transforms LLM agents from text-based assistants into programmable operators of domain-specific workflows. Developing a customized MCP server involves exposing existing computational functionalities as standardized endpoints that an LLM can discover and invoke. At its core, an MCP server acts as a middleware layer: it listens to structured requests from an LLM, executes the corresponding tool or script, and returns results in a machine-readable format. This abstraction allows LLMs to interact with complex software pipelines without directly managing their low-level commands.

\begin{figure}[!hbt]
\centering
\includegraphics[width=0.45\linewidth]{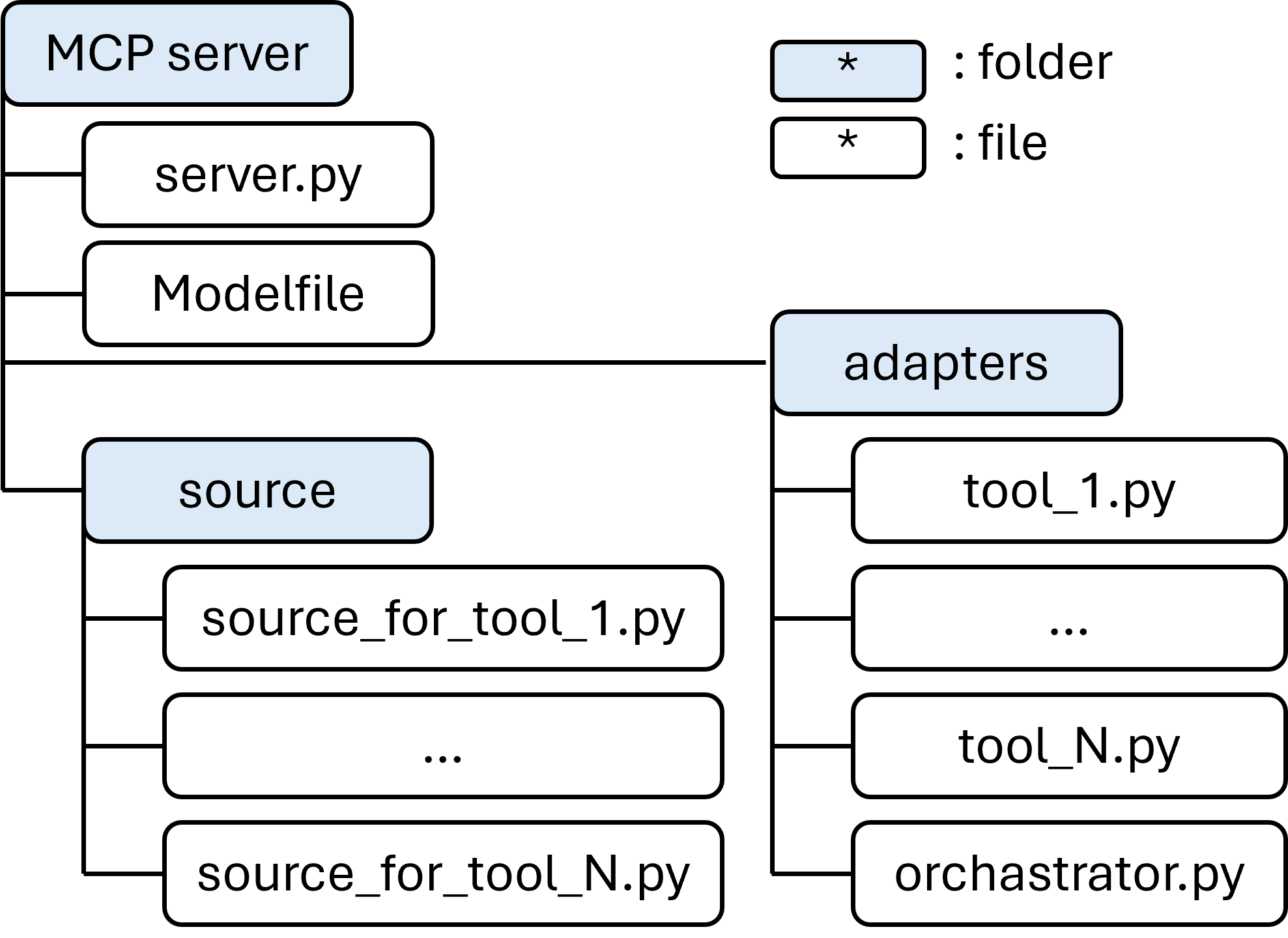}
\caption{Architecture of a customized Model Context Protocol (MCP) server. The server exposes computational tools through a standardized interface. Each tool can be implemented through adapters that can call functions in the source folder with correct configurations.  }
\label{fig:mcp_architecture}
\end{figure}

A typical MCP server project follows a modular architecture (visually illustrated in \ref{fig:mcp_architecture}):  

\begin{itemize}
    \item \textbf{Server layer} (\texttt{server.py}): It launches the server, registers tools, and defines available endpoints. When the tools are registered, detailed contexts for each tool should be provided.  
    \item \textbf{Modelfile}: This file contains system prompts that contain foundational instructions for an AI model's behavior, capabilities, and operational constraints before any user interaction occurs.
    \item \textbf{Adapter layer} (\texttt{adapters/}): This is where the real work happens. It contains specialized modules (\texttt{i.e., tool\_N.py}) that know how to interact with different engineering tools (\texttt{i.e., source\_for\_tool\_N.py}). Each tool is annotated with metadata that describes its purpose and input/output requirements. This folder also contains the \texttt{orchestrator.py} tool, which serves as a central reasoning agent that can plan the usage of multiple tools in this adapter layer from the user query and manage their execution. 
    \item \textbf{Source scripts} (\texttt{source/}): This folder contains real CAE-related functions (\texttt{i.e., source\_for\_tool\_N.py}) that are hard-coded by human experts or other LLMs. The \texttt{tool\_N.py} modules in the adapter folder configure input arguments from the user prompt and call the correct functions in this folder.
\end{itemize}

This modular design supports scalability, that is, new tools can be added incrementally without disrupting existing functionality, and engineers can tailor the server to specific CAE tasks. By encapsulating domain expertise within MCP tools, the server effectively becomes a knowledge repository that LLMs can autonomously explore and apply to new problems.

\section{C-HiDeNN-TD approximation}
\label{A:chidenn}
Leveraging the universal approximation theorem, MLP has been widely adopted as a global basis function in deep learning-based solvers \cite{raissi2019physics}. However, MLP exhibits potential caveats when approximating PDE solutions. For example, MLP enforces initial/boundary conditions indirectly by using a penalty term in the loss function, an approach that cannot guarantee that the conditions will be met exactly. Moreover, despite its differentiable nature, the numerical integration of MLP is not straightforward and is more expensive than standard Gaussian integration. As a result, most MLP PDE solvers use collocation methods and rely on stochastic optimization methods to minimize the PDE residual. This means that the convergence and stability of the solution are not guaranteed, leading to potential unreliability.

To overcome these potential caveats, we leverage a novel AI-enhanced basis function called the Convolution Hierarchical Deep-learning Neural Network (C-HiDeNN). This approach marries the merits of both locally supported finite element shape functions and the flexibility of machine learning. Consequently, C-HiDeNN maintains all the essential properties of finite element approximation, such as the Kronecker delta and the partition of unity \cite{park2023convolution}. Numerical integration of C-HiDeNN basis functions is also straightforward with Gaussian quadrature. As a result, the weak form of the PDE can be adopted similarly to the standard finite element.

\begin{figure}[!hbt]
\centering
\includegraphics[width=0.8\linewidth]{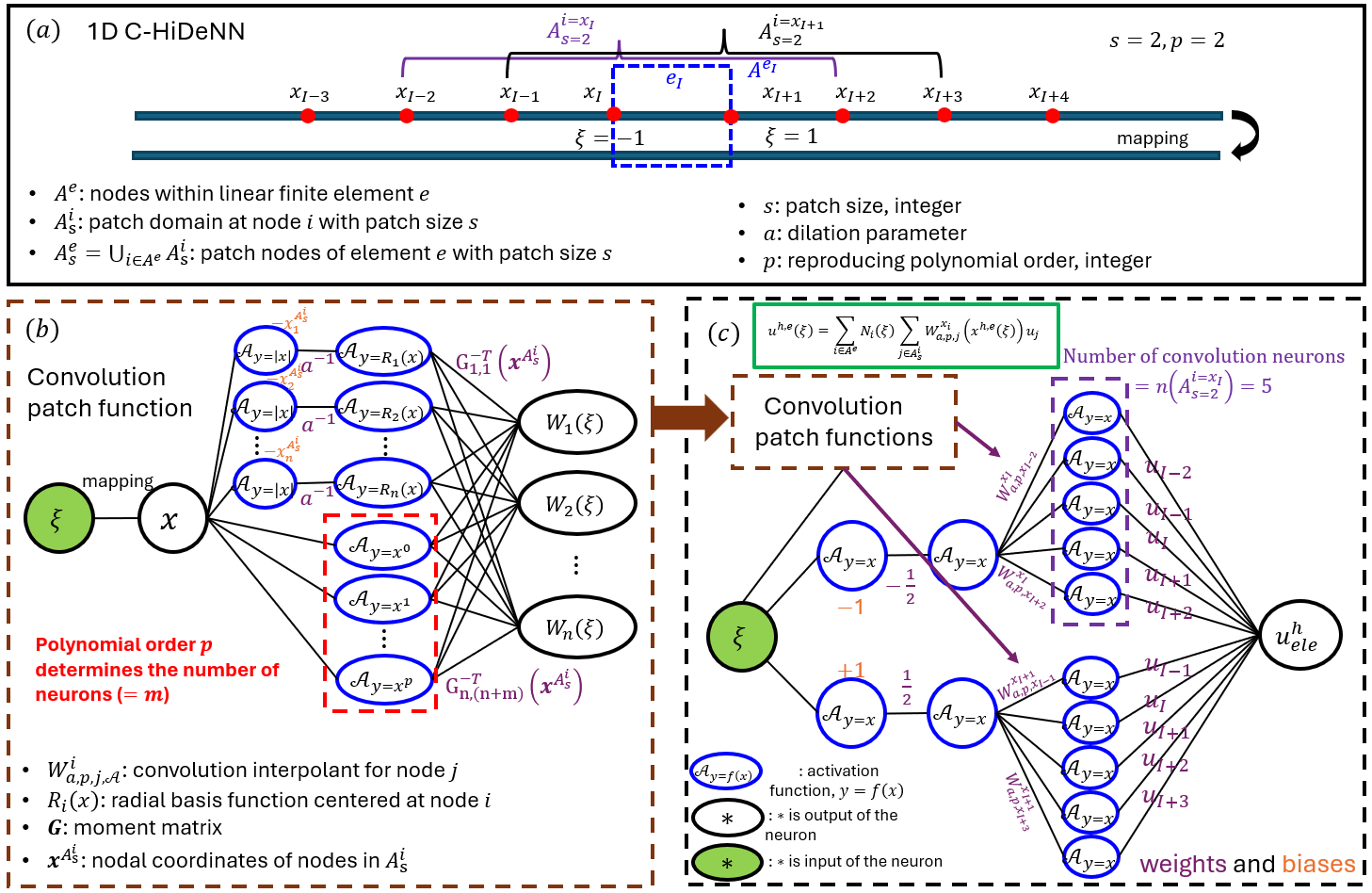}
\caption{(a) Covolution patch in 1D C-HiDeNN shape (basis) function (b) Construction of convolution patch function (c) C-HiDeNN shape function as MLP with 3 hidden layers, adapted from  \cite{park2023convolution}}
\label{shape_fun}
\end{figure}

The detailed formulation of the C-HiDeNN basis function is shown in \ref{shape_fun}. With C-HiDeNN, a 1D scalar field $u(x)$ defined in each element within a domain $\Omega_{x}$ can be approximated using the C-HiDeNN interpolation as:
\begin{equation}
    u_{e}^{h}(x)=\sum_{i\in A^{e}}N_{i}(x)\sum_{j\in A_{s}^{i}}\mathcal{W}_{ \bm{x^*},a,s,p,j,\mathcal{A}}^{i}(x)u_{j}=\sum_{k\in A_{s}^{e}}\widetilde{N}_{k}(x;\bm{x^*},a,s,p,\mathcal{A})u_{k}
\end{equation}
where $u_j$  is the nodal value and $u_j=u(x_j)$; $N_i(x)$ is the linear finite element shape function at node $i$; $\mathcal{W}_{\bm{x^*},a,s,p,j,\mathcal{A}}^i$ is the convolution patch function at node $j$ for the $i$-th nodal patch, which can be represented as a partially connected (pruned) MLP, as illustrated in Fig. \ref{shape_fun}(b). Note that the weights and biases of the ``pruned neural network" are replaced by the nodal positions $\bm{x^*}$ and the hyperparameters. The convolution patch functions are controlled by the activation function $\mathcal{A}$ and three hyperparameters: patch size $s$ which controls nodal connectivity; the dilation parameter $a$ which normalizes distances between patch nodes; and reproducing order $p$ which defines the types/orders of activation functions to be reproduced by the patch functions. As a result, the univariate function can be approximated as:

\begin{equation}
    u^{h}(x)=\sum_{k=1} ^ {nnode}\widetilde{N}_{k}(x;\bm{x^*},a_{k},s_{k},p_{k}, \mathcal{A}_k)u_{k}
\label{full_chidenn}
\end{equation}

\noindent
where $nnode$ is the total number of nodes and $k$ is the nodal index. It should be noted that hyperparameters $\bm{x^*}, a,s,p$, and activation $\mathcal{A}$ can vary between different nodal patches since C-HiDeNN can optimize these hyperparameters such as machine learning parameters, rendering an adaptable functional space without altering the number of global nodes or hidden layers. 

For ease of discussion, in the remaining part, we fix the hyperparameters $\bm{x^*},a_{k},s_{k},p_{k}$ and the activation function $\mathcal{A}_k$ in C-HiDeNN, and omit them in the notation. As a result, we obtain the C-HiDeNN-TD approximation of a univariate function.

\begin{equation}
    u^{h}(x)=\sum_{k=1} ^ {nnode}\widetilde{N}_{k}(x)u_{k}
\label{td_eq}
\end{equation}

\section{Detailed mathematical derivation of TAPS}
\label{app:A}
In this Appendix, we show the detailed mathematical derivation of TAPS for a 1D space, 1D parameter and 1D time problem. The derivation of this problem serves as the detailed template for the mathematical derivation of other more complex parametric PDEs. The parametric transient heat transfer equation is shown below.
\begin{equation}
    \frac{\partial u}{\partial t}-\frac{\partial }{\partial x}\left( \alpha \frac{\partial u}{\partial x}\right)=f(x, \alpha,t)
\label{A:pde_eqn1}
\end{equation}
subject to homogeneous boundary conditions and initial conditions. This equation has 3 independent variables ($D=3$), i.e., spatial variable $x$, parametric variable $\alpha$ and temporal variable $t$. The S-P-T Galerkin weak form of this problem can be written as follows:

\begin{equation}
    \int_{\Omega}\delta u \nabla_{t} u d\Omega +\int_{\Omega}\nabla_{x}\delta u\cdot \alpha\nabla_{x}ud\Omega - \int_{\Omega}\delta u f d\Omega = 0
\label{galerkin1}
\end{equation}
The trial function (solution) is approximated using the C-HiDeNN-TD approximation.

\begin{equation}
    u(x,k,t)=\sum_{m=1}^M u_{x}^{(m)}(x)u_{\alpha}^{(m)}(\alpha)u_{t}^{(m)}(t)
\label{td1}
\end{equation}
The corresponding test function is obtained using the variational principle.

\begin{equation}
    \delta u(x,\alpha,t)=\underbrace{\sum_{m=1}^M\delta u_{x}^{(m)}(x)u_{\alpha}^{(m)}(\alpha)u_{t}^{(m)}(t)}_\textrm{spatial variation}+\underbrace{\sum_{m=1}^M u_{x}^{(m)}(x)\delta u_{\alpha}^{(m)}(\alpha)u_{t}^{(m)}(t)}_\textrm{parametric variation}+\underbrace{\sum_{m=1}^Mu_{x}^{(m)}(x)u_{\alpha}^{(m)}(\alpha)\delta u_{t}^{(m)}(t)}_\textrm{temporal variation}
\label{test1}
\end{equation}
As shown in Eq. \ref{test1}, the test function consists of 3 variational terms: spatial, parametric, and temporal variations. Each of the variations will contribute to the final matrix form in each subspace iteration. As an example, we first plug Eq. \ref{td1} and the spatial variation term of Eq. \ref{test1} into Eq. \ref{galerkin1} to obtain the S-P-T weak form terms corresponding to spatial variation:

\begin{equation}
    \begin{gathered}\underbrace{\int_{\Omega}\sum_{m=1}^M\sum_{n=1}^M \left[\nabla\delta u_{x}^{(m)}({x}) \nabla u_{x}^{(n)}({x})d{x}\right]\cdot \left[u_{\alpha}^{(m)}(\alpha) \alpha u_{\alpha}^{(n)}(\alpha)d{\alpha}\right]\cdot \left[u_t^{(m)}({t})u_{t}^{(n)}({t})d{t}\right]}_\textrm{diffusion term}+\\\underbrace{\int_{\Omega}\sum_{m=1}^{M}\sum_{n=1}^{M}\left[\delta u_{x}^{(m)}({x})u_{x}^{(n)}(x)d{x}\right]\cdot \left[u_{\alpha}^{(m)}({\alpha})u_{\alpha}^{(n)}({\alpha})d{\alpha}\right]\cdot  \left[u_{t}^{(m)}({t}) 
 \nabla_{t}u_{t}^{(n)}({t})dt\right]}_\textrm{time derivative term} - \\\underbrace{\int_{\Omega}\sum_{m=1}^M\left[\delta u_{x}^{(m)}({x})f(x)d{x}\right]\cdot \left[u_{\alpha}^{(m)}({\alpha})d{\alpha}\right]\cdot \left[u_{t}^{(m)}({t})d{t}\right]}_\textrm{forcing term}\end{gathered}
\label{subspace_x_expand_simple}
\end{equation}
1D C-HiDeNN shape functions are used to approximate each univariate function: 

\begin{eqnarray}
 u_d^{(n)}(x_d) = \widetilde{N}^{[d]}_{n_d'}(x_d)u_{n_d'n}^{[d]} \quad  (\text{no sum on $d$}) \nonumber\\ 
 \delta u_d^{(m)}(x_d) = \widetilde{N}^{[d]}_{n_d}(x_d)\delta u_{n_dm}^{[d]} \quad   (\text{no sum on $d$}) 
\label{shape}
\end{eqnarray}
Here we use Einstein summation to simply the notation. The free index $d$ refers to dimension and $d=x,k$ or $t$. The gradient of the interpolated variable can be computed using the shape function derivative $\widetilde{B}^{[d]}_{n_d}(x_d) = \frac{d \widetilde{N}^{[d]}_{n_d}(x_d)}{d x_d}$.

\begin{eqnarray}
 \nabla_{x_d}u_d^{(n)}(x_d) =   \widetilde{B}^{[d]}_{n_d'}(x_d)  u_{n_d'n}^{[d]} \quad   (\text{no sum on $d$}) \nonumber\\ 
 \nabla_{x_d}\delta u_d^{(m)}(x_d) =  \widetilde{B}^{[d]}_{n_d}(x_d)  \delta u_{n_dm}^{[d]}   \quad   (\text{no sum on $d$})
\label{shape_deri}
\end{eqnarray}
Plugging Eqs. \ref{shape} - \ref{shape_deri} into Eq. \ref{subspace_x_expand_simple}, the diffusion term in Eq. \ref{subspace_x_expand_simple} can be rewritten as: 

\begin{equation}
\sum_{m=1}^M\sum_{n=1}^M \underbrace{\int_{\Omega_x} \widetilde{B}_{n_x}(x)  \delta u_{n_xm}^{[x]} \widetilde{B}_{n_x'}(x)  u_{n_x'n}^{[x]} dx}_\textrm{spatial term}\cdot \underbrace{\int_{\Omega_\alpha}\widetilde{N}_{n_\alpha}(\alpha)u_{n_\alpha m}^{[\alpha]} \alpha\widetilde{N}_{n_\alpha'}(\alpha)u_{n_\alpha'n}^{[\alpha]}d{\alpha}}_\textrm{parametric term}\cdot \underbrace{\int_{\Omega_t}\widetilde{N}_{n_t}(t)u_{n_tm}^{[t]}\widetilde{N}_{n_t'}(t)u_{n_t'n}^{[t]}d{t}}_\textrm{temporal term}
\label{subspace_x_expand_diffusion}
\end{equation}
As can be readily seen from Eq. \ref{subspace_x_expand_diffusion}, after doing 1D integration of each term, the parametric and temporal terms can be treated as coefficient matrices:

\begin{eqnarray}
 C^{[\alpha]}_{mn} = \int_{\Omega_\alpha}\widetilde{N}_{n_\alpha}(\alpha)u_{n_\alpha m}^{[\alpha]} \alpha\widetilde{N}_{n_\alpha'}(\alpha)u_{n_\alpha'n}^{[\alpha]}d{\alpha} \nonumber\\ 
 C^{[t]}_{mn} =  \int_{\Omega_t}\widetilde{N}_{n_t}(t)u_{n_tm}^{[t]}\widetilde{N}_{n_t'}(t)u_{n_t'n}^{[t]}d{t}
\label{eq:coef}
\end{eqnarray}
as the only free indices are $m$ and $n$. Substituting the coefficient matrices and rearranging different terms in Eq. \ref{subspace_x_expand_diffusion}, we have:

\begin{equation}
\sum_{m=1}^M \delta u_{n_xm}^{[x]} \sum_{n=1}^M  \left[\int_{\Omega_x}\widetilde{B}_{n_x}(x) \widetilde{B}_{n_x'}(x) dx\right]  \cdot C^{[\alpha]}_{mn} C^{[t]}_{mn} \cdot u_{n_x'n}^{[x]}
\label{subspace_x_expand_diffusion1}
\end{equation}
Like standard FEM, we can define $\int_{x}\widetilde{B}_{n_x}(x) \widetilde{B}_{n_x'}(x) dx$ as the 1D stiffness matrix $K^{[x]}_{n_xn_x'}$ of $x$ dimension in Eq. \ref{subspace_x_expand_diffusion1}. Furthermore, the following fourth-order tensor can be defined (no sum on $m, n$):

\begin{equation}
A_{n_xn_x'mn}^{[x]} =  K^{[x]}_{n_xn_x'} C^{[\alpha]}_{mn} C^{[t]}_{mn}
\label{a_tensor}
\end{equation}
Therefore, Eq. \ref{subspace_x_expand_diffusion} can be further simplified as follows:
\begin{equation}
 \delta u_{n_xm}^{[x]}  A_{n_xn_x'mn}^{[x]} u_{n_x'n}^{[x]}
\label{subspace_x_expand_diffusion2}
\end{equation}
The 4-th order tensor $A_{n_xn_x'mn}^{[x]}$ can be reshaped as a 2nd order tensor $\mathbb{A}_{IJ}^{[x]}$. The trial and test function solution vectors can also be vectorized: 

\begin{subequations}
\begin{gather}
    A_{n_xn_x'mn}^{[x]} =  \mathbb{A}_{IJ}^{[x]}\\
 \text{vec}(\delta \mathbb{U}^{[x]})_I = \left[\text{vec}\left(\delta u_{n_xm}^{[x]}\right)\right]_I \nonumber\\ 
  \text{vec}(\mathbb{U}^{[x]})_J = \left[\text{vec}\left(u_{n_x'n}^{[x]}\right)\right]_J
\end{gather}
\end{subequations}
As a result, Eq. \ref{subspace_x_expand_diffusion2} can be rewritten in the following matrix form:

\begin{equation}
\text{vec}(\delta\mathbb{U}^{[x]})^T  \mathbb{A}^{[x]} \text{vec}(\mathbb{U}^{[x]})
\label{subspace_x_expand_diffusion3}
\end{equation}
Following the same procedure, we can obtain matrix forms corresponding to the time derivative term $\text{vec}(\delta\mathbb{U}^{[x]})^T  \mathbb{B}^{[x]} \text{vec}(\mathbb{U}^{[x]})$, and the forcing term $\text{vec}(\delta\mathbb{U}^{[x]})^T\text{vec}(\mathbb{Q}^{[x]})$ in the spatial variational part of Eq. \ref{subspace_x_expand_simple}. Similar structures can also be obtained for the parametric and temporal variational parts of the test function in Eq. \ref{test1}. Denoting $\mathbb{K}^{[d]} = \mathbb{A}^{[d]} + \mathbb{B}^{[d]}$, the matrix form of the generalized S-P-T Galerkin form in Eq. \ref{galerkin1} can be written as:

\begin{subequations}
\begin{gather*}
\underbrace{\text{vec}(\delta\mathbb{U}^{[x]})^T\mathbb{K}^{[x]}\text{vec}(\mathbb{U}^{[x]})- \text{vec}(\delta\mathbb{U}^{[x]})^T\text{vec}(\mathbb{Q}^{[x]})}_\textrm{spatial variational part} + \underbrace{\text{vec}(\delta\mathbb{U}^{[\alpha]})^T\mathbb{K}^{[\alpha]}\text{vec}(\mathbb{U}^{[\alpha]}) - \text{vec}(\delta\mathbb{U}^{[\alpha]})^T\text{vec}(\mathbb{Q}^{[\alpha]})}_\textrm{parametric variational part} \\ +  \underbrace{\text{vec}(\delta\mathbb{U}^{[t]})^T\mathbb{K}^{[t]}\text{vec}(\mathbb{U}^{[t]}) -
\text{vec}(\delta\mathbb{U}^{[t]})^T\text{vec}(\mathbb{Q}^{[t]})}_\textrm{temporal variational part}   = 0
\end{gather*}
\label{part1}
\end{subequations}
The above equation is a nonlinear system of equations since $\mathbb{K}^{[d]}$ depends on solution vectors in other dimensions. Using subspace iteration, this nonlinear system can be recast into a series of linear systems. For example, in the subspace iteration in the $x$ direction, we let the variation in other dimensions, i.e.,  $\text{vec}(\delta\mathbb{U}^{[\alpha]})$ and $\text{vec}(\delta\mathbb{U}^{[t]})$ equal to 0. Due to the arbitrariness of the test function vector $\text{vec}(\delta\mathbb{U}^{[x]})$,  the above equation becomes:
\begin{equation}
\mathbb{K}^{[x]}\text{vec}(\mathbb{U}^{[x]})- \text{vec}(\mathbb{Q}^{[x]}) = 0
\end{equation}
After $\text{vec}(\mathbb{U}^{[x]})$ is obtained, we update the matrix $\mathbb{K}^{[\alpha]}$ and the subspace iteration in the $\alpha$ dimension can be written as:
\begin{equation}
\mathbb{K}^{[\alpha]}\text{vec}(\mathbb{U}^{[\alpha]})- \text{vec}(\mathbb{Q}^{[\alpha]}) = 0
\end{equation}
We keep iterating in different dimensions until the variation of the solution vector in each dimension is within the tolerance.

\section{Prompt for mathematical derivation of TAPS solver}
\label{app:prompt}

\begin{figure}[!hbt]
\centering
\includegraphics[width=0.9\linewidth]{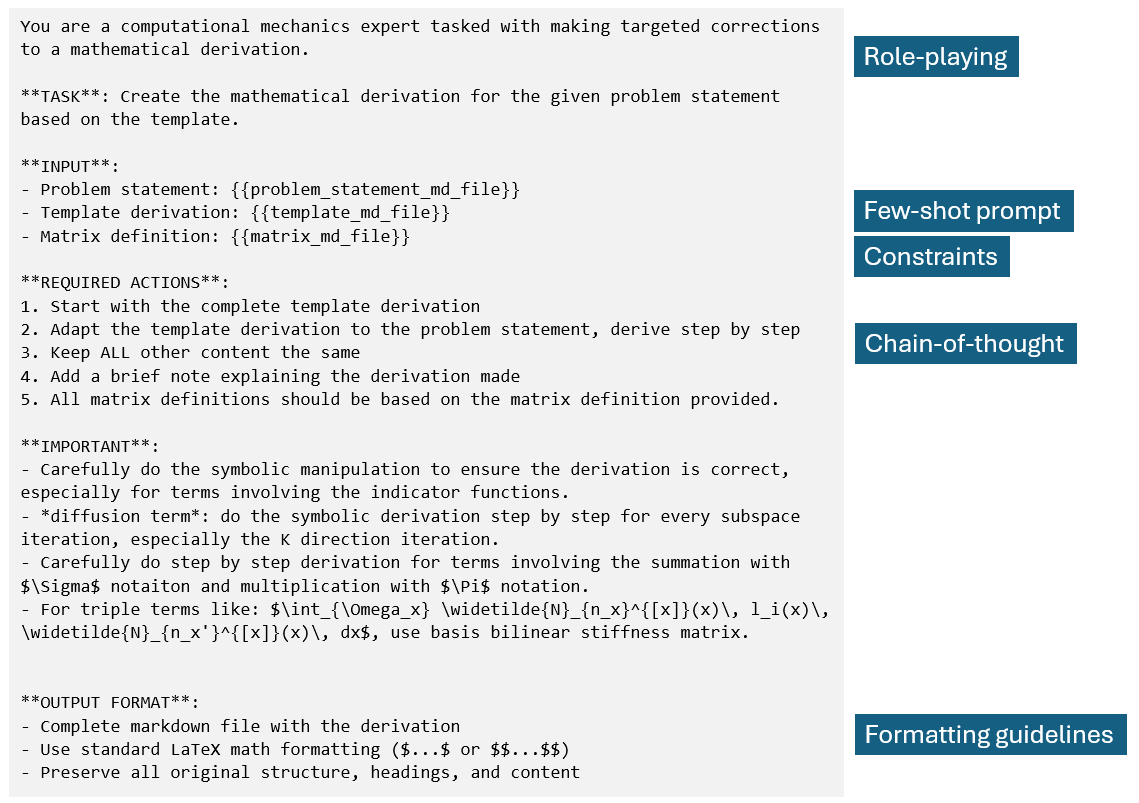}
\caption{Prompt used for mathematical reasoning to derive the discretized form of different PDEs in the TAPS framework.}
\label{mathPrompt}
\end{figure}

In this appendix, we present the complete prompt for the mathematical derivation of new given PDEs for the TAPS solver. As shown in Fig. \ref{mathPrompt}, it consists of 5 major parts: 1. role-playing, where we set up context for the LLMs; 2. few-shot prompt, where the template mathematical derivation is provided; 3. constraints, by which we apply the matrices that the derivation can use; 4. chain-of-thought, where we enforce LLMs to derive the equations step-by-step; 5. formatting guidelines, by which we instruct LLMs to generate a well-formatted markdown with all equations written in \LaTeX.

\bibliography{reference}

\end{document}